\documentclass[12pt]{article}
\usepackage[utf8]{inputenc}
\usepackage{geometry}
\usepackage{graphicx}
\usepackage{array}
\usepackage{subfig}
\usepackage{slashed}
\usepackage{amsmath}
\usepackage{amsfonts}
\usepackage{amssymb}
\usepackage[cm]{fullpage}
\usepackage{cite}
\usepackage{epstopdf}
\usepackage{comment}
\usepackage{hyperref}
\usepackage{cancel}
\usepackage{mathrsfs}

\usepackage{cleveref}
\crefname{section}{§}{§§}
\Crefname{section}{§}{§§}

\usepackage{booktabs} 

\numberwithin{equation}{section}

\def\0{{(0)}}
\def\1{{(1)}}
\def\2{{(2)}}

\def\<{\langle }
\def\>{\rangle }

\newcommand{\bea}{\begin{eqnarray}}

\newcommand{\eea}{\end{eqnarray}}

\newcommand{\be}{\begin{equation}}
\newcommand{\ee}{\end{equation}}
\newcommand{\ba}{\begin{align}}
\newcommand{\ea}{\end{align}}

   \makeatletter
  \let\over=\@@over \let\overwithdelims=\@@overwithdelims
  \let\atop=\@@atop \let\atopwithdelims=\@@atopwithdelims
  \let\above=\@@above \let\abovewithdelims=\@@abovewithdelims
\renewcommand\section{\@startsection {section}{1}{\z@}%
                                   {-3.5ex \@plus -1ex \@minus -.2ex}
                                   {2.3ex \@plus.2ex}%
                                   {\normalfont\large\bfseries}}

\renewcommand\subsection{\@startsection{subsection}{2}{\z@}%
                                     {-3.25ex\@plus -1ex \@minus -.2ex}%
                                     {1.5ex \@plus .2ex}%
                                     {\normalfont\bfseries}}

\newcommand{\beq}{\begin{equation}}
\newcommand{\eeq}{\end{equation}}
\newcommand{\beqa}{\begin{eqnarray}}
\newcommand{\eeqa}{\end{eqnarray}}
\newcommand{\beqar}{\begin{eqnarray*}}

\def\[{\[}
\def\]{\]}

\newcommand{\bd}[1]{\begin{fmffile}{#1}\begin{fmfgraph*}}
\newcommand{\ed}{\end{fmfgraph*}\end{fmffile}}

\begin{document}

\begin{titlepage}

\begin{flushright}
{\small  CERN--TH--2018--112},
{\small  MPP--2018--87},
{\small LMU--ASC 29/18}
\end{flushright}

\vspace{0.6cm}

\begin{center}

{\LARGE{\textsc{Spin-four ${\cal N}=7$ W-Supergravity:}}}

\vskip0.4cm

 {\LARGE{\textsc{S-fold and Double Copy Construction}}}

\vspace{1cm}
{\large \bf Sergio Ferrara$^{a,b,c}$,   \, Dieter L\"ust$^{a,d,e}$}

\vspace{1cm}

{\it

\vskip-0.3cm
\centerline{ $^{\textrm{a}}$ CERN, Theory Department,}
\centerline{1211 Geneva 23, Switzerland}
\medskip
\centerline{ $^{\textrm{b}}$ INFN, Laboratori Nazionali di Frascati,}
\centerline{Via Enrico Fermi 40, 00044 Frascati, Italy}
\medskip
\centerline{ $^{\textrm{c}}$ Department of Physics and Astronomy}
\centerline{and Mani L. Bhaumik Institute for Theoretical Physics, U.C.L.A}
\centerline{Los Angeles CA 90095-1547, U.S.A.}
\medskip
\centerline{ $^{\textrm{d}}$ Arnold--Sommerfeld--Center for Theoretical Physics,}
\centerline{Ludwig--Maximilians--Universit\"at, 80333 M\"unchen, Germany}
\medskip
\centerline{$^{\textrm{e}}$ Max--Planck--Institut f\"ur Physik,
Werner--Heisenberg--Institut,}
\centerline{ 80805 M\"unchen, Germany}

}

\end{center}

\vskip0.5cm
\abstract{{In the present investigation we consider the possibility of having new  massive, higher spin W-supergravity theories, which do not exist as four-dimensional
perturbative  models. These theories are based on a double copy construction of two supersymmetric field theories, where at least one factor 
is given by a  ${\cal N}=3$ field theory, which is a non-perturbative S-fold of ${\cal N}=4$ super Yang-Mills theory.
In this way, we can obtain as S-folds a new
${\cal N}=7$ (corresponding to 28 supercharges) W-supergravity  and its ${\cal N}=7$ W-superstring counterpart, which both do not exist as four-dimensional
perturbative  models with  an (effective) Langrangian description.
  The resulting field resp. string theory does not contain any massless states, but instead a massive higher spin-four supermultiplet of the 
${\cal N}=7$ supersymmetry algebra.
Furthermore we also construct  a four-dimensional heterotic S-fold with ${\cal N}=3$ supersymmetry. It again does not exist as perturbative heterotic string model and
can be considered as the heterotic counterpart of the ${\cal N}=3$  superconformal field theories,   which  were  previously constructed in the context of type I orientfold models.
}}

\end{titlepage}

\newpage

\tableofcontents
\break

\section{Introduction}

For space-dimensions $D\geq4$ supergravity theories \cite{Freedman:1976xh,Deser:1976eh}
have been investigated in great detail 
(for overviews see e.g. \cite{sugra,Freedman:2012zz}).
Moreover this discussion has been extended in \cite{deWit:1992psp}, where also a full classification of all possible number of supercharges has been obtained  
for the case of $D=3$.\footnote{Recently, two-dimensional string constructions 
   with exotic supersymmetries were constructed in \cite{Florakis:2017zep}.}
One of the main results is that a supergravity theory with 28 supercharges,  here commonly referred as ${\cal N}=7$ supergravity using four-dimensional language, is not possible.
In fact, when constructing the corresponding superalgebras and also using CPT symmetry,  the local ${\cal N}=7$ supersymmetry is always automatically extended to
the case of maximal supersymmetry, namely to  ${\cal N}=8$ supergravity with 32 supercharges. The absence of a perturbative  ${\cal N}=7$ supergravity theory is analogous to
the absence of  any perturbative, four-dimensional supersymmetric gauge theories with rigid ${\cal N}=3$ supersymmetry.  Again, using same kind of arguments as in supergravity, one can show that the ${\cal N}=3$ superalgebra of a CPT invariant gauge theory action is always extended to the maximal case of ${\cal N}=4$ rigid supersymmetry.
To summarize,  the three main assumptions, which enter the  ${\cal N}=7$ supergravity as well as the ${\cal N}=3$ field theory no-go theorems, are as follows:

\begin{itemize}

\item
The theory is weakly coupled.
\item
There exist a CPT invariant Langrangian description.
\item
For the case of supergravity, there exist a massless, N-extended spin-two supergravity multiplet and no higher spin massless multiplets.

\end{itemize}

The prime goal of this paper is to provide some good arguments for the existence of a new kind of ${\cal N}=7$ locally supersymmetric theory in $D=4$, which is not based on a conventional, perturbative supergravity
action, but which however is strongly coupled and does not exists as a Langrangian theory. It will not contain a massless spin-two ${\cal N}=7$
supergravity multiplet, but  a massive  ${\cal N}=7$ multiplet of higher spin-four, which realizes the  ${\cal N}=7$ supersymmetry algebra. In analogy to higher spin theories in two dimensions
we call these theories ${\cal N}=7$ W-supergravity theories.\footnote{Alternatively W stands for Weyl, as it will be justified later.}
In this way the above assumptions for the no-go theorems  are circumvented.
This new  ${\cal N}=7$ W-supergravity theory is analogous to the  ${\cal N}=3$ supersymmetric gauge theories, which were recently constructed in 
\cite{Garcia-Etxebarria:2015wns,Aharony:2015oyb,Aharony:2016kai} and which are
 non-perturbative in the sense that they only exist at strong gauge coupling and that they also do not allow for a known Lagrangian formulation
with a massless gauge supermultiplet. Furthermore they are conjectured to be superconformal.
The ${\cal N}=3$ supersymmetric gauge theories can be constructed as
  S-folds starting from    ${\cal N}=4 $ supersymmetric Yang-Mills gauge theory and projecting on those operators, which are invariant under
  a combined action of R-symmetry and S-duality:
   \begin{equation}
{\rm Field~theory~S-fold:}\quad ({\cal N}=3_{SYM})\equiv ({\cal N}=4_{SYM})/(R\times S)\, .
 \end{equation}
A closely related, holographically dual  ${\cal N}=6$ supergravity in $AdS_5$ space was already constructed in \cite{Ferrara:1998zt}, with the same kind of S-fold projection
being used as in the ${\cal N}=3$ supersymmetric gauge theories of
\cite{Garcia-Etxebarria:2015wns,Aharony:2015oyb,Aharony:2016kai}.

In the first part of the paper we will use the 
compelling observation that Einstein gravity can be regarded as the square of  two Yang-Mills theries. This is the socalled double copy construction 
of (super)gravity theories \cite{Bern:2008qj,Bern:2010yg,Bern:2010ue}, which was utilized for
\begin{equation}
{\cal N}=8~{\rm Sugra}\equiv({\cal N}=4~{\rm YM})^2
\end{equation}
 to compute scattering amplitudes in ${\cal N}=8$ supergravity from ${\cal N}=4$ super Yang-Mills amplitudes.
The double copy construction
was also used to obtain all possible standard supergravity theories with extended local supersymmetry
\cite{Chiodaroli:2014xia,Anastasiou:2014qba,Chiodaroli:2015rdg,Chiodaroli:2015wal,Borsten:2015pla,Cardoso:2016ngt,Anastasiou:2017nsz,Chiodaroli:2017ehv}.
Here we will construct  new W-supergravities with ${\cal N}=3,4,5,6,7$, namely 
  we will generalize the 
double copy construction by considering tensor products 
with one
factor being the ${\cal N}=3$ supersymmetric gauge theory described above. 
In particular the ${\cal N}=7$ W-supergravity 
will be constructed   as a  double copy of a non-perturbative  theory with a ${\cal N}=4$ super Yang-Mills theory:
 \begin{eqnarray}
{\cal N}=7 ~{\rm W-Sugra}
\equiv({\cal N}=4~{\rm YM})\otimes ({\cal N}=3~{\rm YM})
\, .
\end{eqnarray}
The two factors respectively refer to the spin-two massive supermultiplets of ${\cal N}=4,3$. They are also denoted as Weyl multiplets, since they corresponds to the
multiplets of the gauge fields of corresponding local superconformal algebra.
As we will discuss, the double copy  will contain as lightest multiplet a massive spin-four supermultiplet of the local ${\cal N}=7$ supersymmetry algebra, which arises
as the tensor product of two spin-two Weyl multiplets \cite{Ferrara:1977ij,Bergshoeff:1980is,vanMuiden:2017qsh,deWit:1979dzm,Ferrara:1977mv,deWit:1978pd}
of the ${\cal N}=3$ and ${\cal N}=4$ theories.

The double copy construction 
of gravity theories can be also traced back to the KLT-relations \cite{Kawai:1985xq} in string theory, 
as well as from intriguing relations between open and closed string amplitudes
 \cite{Stieberger:2009hq,Stieberger:2016lng}.
 In fact, in the second part of the paper we propose
 a closely related  construction of new, non-perturbative S-folds in string theory, which we will call W-superstrings.
In fact, as it is well-known from the ''old days'' of perturbative string constructions,  space-time supersymmetry and world-sheet supersymmetry are closely linked together
in string theory.
On the heterotic side, 
four-dimensional perturbative  heterotic string constructions  lead to ${\cal N}$-extended supergravity theories with following number of supercharges: 
\begin{eqnarray}\label{hetN}
{\cal N}_H=1\, ,\quad {\cal N}_H=2\, ,\quad{ \rm and}\quad  {\cal N}_H=4\,,
\end{eqnarray}
Here the space-time supercharges originate from the superconformal, right-moving fermionic string sector of the heterotic string. The bosonic, left-moving sector of the heterotic string however
does not provide any further space-time supercharges.

In type II constructions, the  possible space-time supercharges are  obtained by building  the tensor product of  superconformal left- and  right-moving fermionic strings.
This  leads to several
four-dimensional ${\cal N}$-extended supergravity theories from  perturbative type II string constructions, which are classified as follows:
\begin{eqnarray}
&~&{\cal N}_{II}=1=0_L+1_R\, ,\quad {\cal N}_{II}=2=0_L+2_R\,, \quad {\cal N}_{II}=2'=1_L+1_R\, ,\nonumber\\
&~&{\cal N}_{II}=3=1_L+2_R\,,\quad
{\cal N}_{II}=4=0_L+4_R\,, \quad {\cal N}_{II}=4'=2_L+2_R\,, \nonumber\\
&~&{\cal N}_{II}=5=1_L+4_R\,, \quad{\cal N}_{II}=6=2_L+4_R\, ,\quad{\cal N}_{II}=8=4_L+4_R\, .\label{typeIIsupercharges}
\end{eqnarray}
Explicit constructions of the vertex operators for the type II space-time supercharges 
as products of left- and right-moving pieces and the associated action of the R-symmetries for all these  theories were provided in 
\cite{Lust:1988yp,Ferrara:1989ud}. 

It is clear from eq.(\ref{hetN}) that on the heterotic side it is perturbatively not possible to construct a  ${\cal N}_H=3$ supergravity theory. This is a reflection of the fact that there is also no 
 perturbative  ${\cal N}=3$ supersymmetric gauge theory. 
 However among the possible type II cases in eq.(\ref{typeIIsupercharges}), there exists  a four-dimensional, perturbative  ${\cal N}_{II}=3$  supergravity theory. Explicit 
 left-right asymmetric type II constructions in particular with ${\cal N}_{II}=3,5,6$ were provided in \cite{Ferrara:1989nm}.
 But we also recognize in eq.(\ref{typeIIsupercharges}) that there is no possible ${\cal N}_{II}=7$ supergravity theory (i.e. with 28 supercharges) from a  perturbative type II constructions.  Therefore
 there does not  exist a standard ${\cal N}=7$ superstring theory with a massless spin-two supermultiplet.
 
In second part of the paper, 
we will  construct particular string theory quotients, called string theory S-folds,
 of four-dimensional heterotic and type II strings by dividing with particular elements of the T- and the S-duality groups.  
 Namely, 
 when constructing the string theory S-fold, we will realize that the R-symmetry in field theory is given in terms of certain T-duality transformations. They act as
 asymmetric rotations on the internal world-sheet coordinates.
 It follows that
  the internal six-dimensional space of the four-dimensional string theory S-folds is a non-geometric space, basically given in terms of a particular asymmetric orbifold
(see e.g. \cite{Condeescu:2012sp,Condeescu:2013yma}).
Related type string models in two dimensions with 28 chiral supercharges  were recently constructed in \cite{Florakis:2017zep}.
In addition we { propose a method} how the field theory S-duality is implemented into the string construction.
The existence of
S-duality in four dimensional heterotic string compactifications was first proposed in \cite{Font:1990gx} and further evidence for this conjecture was subsequently given in 
\cite{Sen:1994fa}. As we will discuss, also S-duality will act as asymmetric rotation on the uncompactified coordinates in a particular way.

Then,
following the same strategy as in field theory, we show  that S-folds of heterotic string compactifications lead to  heterotic string theories with ${\cal N}_H=3$ supersymmetry:
 \begin{equation}
 {\rm String~theory~S-fold:}\quad({\cal N}=3_H)\equiv ({\cal N}=4_H)/(T\times S)\, .
 \end{equation}
Moreover, by building the tensor product of  a left-moving fermionic string with ${\cal N}_L=4$ together with an right moving  fermionic S-fold with ${\cal N}_R=3$
  we will obtain a type II S-fold with ${\cal N}_{II}=7$ supersymmetry:
 \begin{eqnarray}
{\rm String~theory~S-fold:}\quad({\cal N}=7_{II})\equiv 
({\cal N}=4)_L\otimes ({\cal N}=3)_R
\, .
\end{eqnarray}
This ${\cal N}=7$ W-superstring theory is  an entirely higher spin theory without a standard massless spin-2 supermultiplet, and hence it does not possess
a standard Langrangian description in terms of a conventional Einstein action coupled to massless spin=3/2 gravitinos. 
It is natural to conjecture thats its effective description of the lowest massive states is just ${\cal N}=7$ W-supergravity, described above.


\section{Double copy construction of massive   W-supergravities}

\subsection{Double copy constructions}

In the following we will outline the
 double copy construction by taking the product of two four-dimensional supersymmetric field theories QFT$({{\cal N}_L})$ and QFT$({{\cal N}_R})$
 with  ${\cal N}_L$ resp. with   ${\cal N}_L$  rigid supersymmetry. This leads to  four-dimensional supergravity theories with ${\cal N}_L+{\cal N}_R$ extended, local supersymmetry:
\begin{equation}  
     {\rm QFT}({{\cal N}_L})  \otimes     {\rm  QFT}({{\cal N}_R})\quad=\quad
                                          {\rm Sugra}({\cal N}_L+{\cal N}_R)\, .
                                          \end{equation}

In order to obtain the spectrum of the double copy, one first has to  determine the operators $\Phi_L$ and $\Phi_R$ with lowest scaling dimension $h_L$ and $h_R$
in each of the two products. Depending on their scaling dimensions, they correspond either to certain massless or massive 
superfields with masses $m_L$ and $m_R$.
In order to get the lowest fields in the double copy, we build the tensor product field
\begin{equation}
\Phi_{L+R}=\Phi_L \otimes \Phi_R , 
\end{equation}
where 
one has to demand that
$h_L = h_R$, i.e. $m_L = m_R$. This condition corresponds to the level matching constraint in string theory.

\subsubsection{Standard massless supergravities as double copy constructions:}

\vskip0.4cm
\noindent
First, we can list  the nine familiar examples of double copy constructions providing standard extended supergravity theories with massless spin-two 
supermultiplets   \cite{Ferrara:1976iq,Freedman:1976nf}.\footnote{The ${\cal N}\geq 4$ and ${\cal N}<4$ cases of these nine double copy constructions where first systematically obtained at the quantum level in  \cite{Bern:2011rj,Carrasco:2012ca}.}
In all these examples the lowest operators of  QFT$({{\cal N}_L})$ and QFT$({{\cal N}_R})$ are the  spin-one vector multiplets $V$, which correspond  to the massless 
$U(1)$ super  multiplets in each of the two field theories.
The double copy is provided by the tensor product of two massless vector multiplets:
  \begin{equation}
{\rm Supergravity}:\quad {\rm Massless ~Spin(2)}=V_{L, {\cal N}_L} \otimes V_{R,{\cal N}_R} \, .
\end{equation} 
The massless spin-two supergravity multiplet is always present in the tensor product.

\vskip0.2cm
\noindent
(i) ${\cal N}_L={\cal N}_R =4$. This leads to the well-known  extendend ${\cal N}=8$ supergravity  in the double copy. The lowest operators in each factor are the massless spin-1 vector multiplets (singleton of ${\cal N}=4$) with 8 bosonic (B) +  8 fermionic (F) massless states.
     Their tensor product  is the massless spin-2 gravity multiplet of ${\cal N}=8$ with $n_B=n_F=128$ massless states.

\vskip0.2cm
\noindent
(ii) ${\cal N}_L=2$, ${\cal N}_R=4$. This leads to standard massless pure ${\cal N}=6$ supergravity.

\vskip0.2cm
\noindent
(iii) ${\cal N}_L=1$, ${\cal N}_R=4$. This leads to standard massless pure ${\cal N}=5$ supergravity.

\vskip0.2cm
\noindent
(iv) ${\cal N}_L=0$, ${\cal N}_R=4$. This leads to standard massless pure ${\cal N}=4$ supergravity.

\vskip0.2cm
\noindent
(v) ${\cal N}_L={\cal N}_R=2$. This leads to massless ${\cal N}=4$ supergravity with two  additional massless ${\cal N}=4$ vector multiplets.

\vskip0.2cm
\noindent
(vi) ${\cal N}_L=1$, ${\cal N}_R=2$. This leads to massless ${\cal N}=3$ supergravity with one additional massless ${\cal N}=3$ vector multiplet.

\vskip0.2cm
\noindent
(vii) ${\cal N}_L=0$, ${\cal N}_R=2$. This leads to massless ${\cal N}=2$ supergravity with one additional massless ${\cal N}=2$ vector multiplet.

\vskip0.2cm
\noindent
(viii) ${\cal N}_L={\cal N}_R=1$. This leads to massless ${\cal N}=2$ supergravity with one additional massless ${\cal N}=2$ hyper multiplet.

\vskip0.2cm
\noindent
(ix) ${\cal N}_L=0$, ${\cal N}_R=1$. This leads to massless ${\cal N}=1$ supergravity with one additional massless ${\cal N}=1$ chiral multiplet.

\vskip0.2cm
\noindent

Note that comparing with the corresponding string constructions, heterotic models can only be obtained if ${\cal N}_L=0$.
Furthermore the only way to obtain massless ${\cal N}=3$ supergravity with the double copy method is as in (vi), namely ${\cal N}_L=1$, ${\cal N}_R=2$.
This model was obtained as a type II string construction in \cite{Ferrara:1989nm} when it was observed that the three complex scalars parametrize the $CP_3$ manifold 
$SU(1,3)/(U(1)\times SU(3))$ in contrast to the ${\cal N}=4$ case where the six real scalars would parametrize $O(1,6)/SO(6)$. Since the vector $\otimes $ vector tensor product 
always produces two scalars, it is more difficult to construct pure ${\cal N}=1,2,3$ supergravities as double copies.\footnote{A way to construct these theories was demonstrated in
\cite{Johansson:2014zca,Johansson:2017bfl}.}
Viceversa pure supergravities exist for ${\cal N}\geq 4$
with ${\cal N}_L=k$, ${\cal N}_R=4$, $k=0,1,2,4$.
The numbers of states of ${\cal N}=k+4$ supergravity is  $16\times 2^{k+1}$ for CPT not self conjugate theories ($k=0,1,2$)
and $16\times 16$ for the ${\cal N}=8$ ($k=4$) CPT self conjugate one.

\subsection{4D massive supermultiplets of ${\cal N}$-extended supersymmetry:}

\vskip0.4cm
\noindent
Before we proceed to consider and to construct  explicit examples of W-supergravity theories, let us provide a general overview about the structure  of massive supermultiplets 
and tensor products between them.\footnote{For massive representations we mean long multiplets
without central charges.} General massive multiplets of ${\cal N}$-extended 4D supersymmetry   \cite{Ferrara:1980ra}
with top spin $j_{max}=j+{{\cal N}\over 2}$ are obtained by tensoring the smallest multiplet with top spin ${{\cal N}/2}$ with a spin-j representation of $SU(2)$. The number of states is
then
\begin{equation}
n_B+n_F=2^{2{\cal N}}(2j+1)\, ,\quad(n_B=n_F)\, .
\end{equation}
For $j=0$ the state of spin$({{\cal N}\over 2}-{k\over 2})$ is the k-fold antisymmetric (traceless) irreducible representation of $USp(2{\cal N})$.

If we tensor two massive states of ${\cal N}_1$ and ${\cal N}_2$ extended supersymmetry, we get a massive multiplet of ${\cal N}_1+{\cal N}_2$ extended 
supersymmetry with total multiplicity
\begin{equation}
S=n_B+n_F=2^{2({\cal N}_1+{\cal N}_2)}
(2j_1+1)(2j_2+1)\, ,\quad(n_B=n_F)\, .
\end{equation}
This representation is reducible into
\begin{equation}
S=\sum_{j=|j_1-j_2|}^{j_1+j_2}(2j+1)2^{2({\cal N}_1+{\cal N}_2)}
\end{equation}
representations. So the representation is irreducible only if $j_2=0$ (or $j_1=0$) for which $S=2^{2({\cal N}_1+{\cal N}_2)}
(2j+1)$.

In our context the four-dimensional ${\cal N}$-extended Weyl multiplets \cite{Ferrara:1977ij}
will be of particular importance. They correspond to 4D massive spin-two fields of the
of ${\cal N}$-extended supersymmetry.
The generic  spin-two massive multiplet (Weyl multiplet) is obtained by tensoring the smallest massive representation of ${\cal N}$-extended supersymmetry 
(with $j_{max}={{\cal N}\over 2}$) with spin $j=2-{{\cal N}\over 2}$. The number of states of such spin-two multiplet are therefore $n_B+n_F=(5-{\cal N})2^{2{\cal N}}$.
In particular for ${\cal N}=4$ it is the smallest long massive multiplet.

So let us list  all the possible Weyl supermultiplets, denoted by $W_{ {\cal N}=k}$ ($k=1,2,3,4$), of supersymmetric field theories.
The spin-two massive multiplet of ${\cal N}=4$ is irreducible with $n_B+n_F= 2^8=256$
 with states in $USp(8)$ representations       \cite{Bergshoeff:1980is}:
 \begin{eqnarray}
W_{{\cal N}=4}:~&   
{\rm Spin}(2) + {\underline{8}}\times {\rm Spin} ( 3/2 )  + {\underline{27}}
\times {\rm Spin}(1) + {\underline{48}}\times {\rm Spin} (1/2) + {\underline{42}}\times {\rm Spin} (0)\, .
    \end{eqnarray}  
   The spin-two massive Weyl multiplet of ${\cal N}=3$ is obtained by tensoring the smallest $j_{max}={{\cal N}\over 2}={3\over 2}$ massive multiplet with $j={1\over 2}$, and it is 
irreducible with states in $USp(6)$ representations where $n_B+n_F=2\times 2^6=128$ \cite{vanMuiden:2017qsh}:
        \begin{eqnarray}     
W_{{\cal N}=3}:   
 {\rm Spin}(2) + {\underline{6}}\times {\rm Spin} ( 3/2 )  + ({\underline{14}}+
{\underline{1}})\times {\rm Spin}(1) + ({\underline{14'}}+{\underline{6}})\times {\rm Spin} (1/2) 
+ {\underline{14}}\times {\rm Spin} (0)      \,.
\end{eqnarray}
Finally for ${\cal N}=2,1$ one obtains \cite{deWit:1979dzm,Ferrara:1977mv}
\begin{eqnarray}
W_{{\cal N}=2}:~    {\rm Spin}(2) + {\underline{4}}\times {\rm Spin} ( 3/2 )  + ({\underline{5}}+{\underline{1}})
\times {\rm Spin}(1) + {\underline{4}}\times {\rm Spin} (1/2) +  {\rm Spin} (0)        \, ,
\end{eqnarray}
where $n_B+n_F= 48$.
\begin{eqnarray}
W_{{\cal N}=1}:~    {\rm Spin}(2) + {\underline{2}}\times {\rm Spin} ( 3/2 )  + 
 {\rm Spin}(1)       \, .
\end{eqnarray}
with $n_B+n_F=16$ states.

The field theory realization of Weyl multiplets allows one to write a unique Lagrangian for all conformal supergravities (${\cal N}\leq4$) 
\cite{Kaku:1977rk,Fradkin:1985am,Berkovits:2004jj,Kallosh:2016xnm}.      
In view of the fact that massive spin-two can also be seen as Weyl multiplets, it is instructive to indicate their $U({\cal N})$ R-symmetry quantum numbers. This identification
is achieved by decomposing $USp(2{\cal N})$ into $U({\cal N})$ representations. The states, which are not gauge fields, are auxiliary fields of extended superconformal
supergravity. For example in the ${\cal N}=1$ spin-two multiplet (2,2(3/2),1)  there are the Weyl graviton, the R-symmetry $U(1)$ gauge bosons and two spin-3/2 gauge fields of
$Q$ and $S$ supersymmetry. There are no auxiliary fields in this case. For the higher  cases there are always  2${\cal N}$ gravitini for $Q$ and $S$ supersymmetry.
The $U({\cal N})$ gauge fields must be in the spin-one sector of the massive multiplet.
For ${\cal N}=2,3,4$ the spin-one representations are easily seen to always contain the adjoint representation of $U({\cal N})$. For example  
in ${\cal N}=3$ the spin-one decomposes into 
$ ({\underline{14}}+
{\underline{1}})\rightarrow \underline 8+\underline 3+{\underline{\bar 3}}+ \underline1$ under the  $U(3)$,
where the $ \underline 8+ \underline1$ are the gauge bosons the $U(3)$ R-symmetry.
In ${\cal N} =4$ the ${\underline{27}}$ of $USp(8)$ decomposes into ${\underline {15}}+\underline{6}+{\underline{\bar 6}}$ of $SU(4)$.
Here the ${\underline {15}}$ correspond to the gauge bosons of the $SU(4)$ R-symmetry, since there is no $U(1)$ factor in the gauge symmetry.
The other states correspond to auxiliary fields, whose presence have the double role of completing the spin-two massive multiplet or, in the field theory side, to ensure
the same number of bosonic and fermionic off-shell degrees of freedom in the Weyl multiplet.

\subsection{The ${\cal N}=3$ S-fold}\label{sfoldft}

Let us recall how,  starting from  the ${\cal N}=4$ supersymmetric SYM theories,  how the strongly coupled ${\cal N}=3$ supersymmetric field theory is obtained by a 
projection, which is a combination of a group element of the R-symmetry and the $SL(2,{\mathbb Z})$ S-duality automorphism group \cite{Garcia-Etxebarria:2015wns}.
The 16  supercharges  of four-dimensional
${\cal N}=4$ supersymmetry are denoted as   $Q^{\alpha, A}$ and $Q^{\dot \alpha, \dot A}$.   Here $\alpha=1,2$ and $\dot \alpha=1,2$ are the spinor indices
of the $({\underline 2},{\underline 1})+({\underline 1},{\underline 2})={\underline 2}_s+{\underline 2}_c$ dimensional spinor representations of the four-dimensional Lorentz group $SO(1,3)\equiv SL(2)\times SL(2)$.
Furthermore the indices $A=1,\dots ,4$ and $\dot A=1,\dots ,4$ denote the spinor  and anti-spinor indices of the two inequivalent spinor representations 
${\underline 4}_s$ and ${\underline 4}_c$ of the four-dimensional R-symmetry group $SO(6)$.
More details on the transformation rules of the supercharges are given in the appendix.

Now following \cite{Garcia-Etxebarria:2015wns}, in order to break  ${\cal N}=4$ supersymmetry  down to  ${\cal N}=3$ supersymmetry, 
we will choose a particular ${\mathbb Z}_4$ rotation, which is embedded in the $SO(6)_R$ R-symmetry group and which acts on the    ${\cal N}=4$   supercharges
as follows:
\begin{eqnarray}
{\rm R-symmetry:}\quad &{~}& Q^{\alpha,A}\rightarrow 
e^{-{i\pi(w_1^{ A}+w_2^{ A}+w_3^{A})\over 2}}Q^{\alpha,A}\, , \nonumber\\
&{~}&Q^{\dot\alpha,\dot A}\rightarrow e^{-{i\pi(w_1^{\dot A}+w_2^{\dot A}+w_3^{\dot A})\over 2}}
Q^{\dot \alpha,\dot A}\, , 
\end{eqnarray}
where the $(w_1^A,w_2^A,w_3^A)=(\pm {1\over 2},\pm {1\over 2},\pm {1\over 2})$ (even number of - signs) are  spinor weights of ${\underline 4}_s$ of the $SO(6)_R$
 group, whereas the $(w_1^{\dot A},w_2^{\dot A},w_3^{\dot A})=(\pm {1\over 2},\pm {1\over 2},\pm {1\over 2})$ (odd number of - signs) are  spinor weights of ${\underline 4}_c$ representation.
  The supercharges specifically transform under under the ${\mathbb Z}_4$ R-symmetry rotation as:
\begin{eqnarray}\label{Rcharges}
Q^{\alpha,A}:\quad &{~}& ( {1\over 2}, {1\over 2}, {1\over 2})\rightarrow e^{-{3i\pi\over 4}}( {1\over 2}, {1\over 2}, {1\over 2})\, ,\nonumber\\
&{~}& ( -{1\over 2},- {1\over 2}, {1\over 2})\rightarrow e^{{i\pi\over 4}}( -{1\over 2},- {1\over 2}, {1\over 2})\, ,\nonumber \\
&{~}& ( -{1\over 2}, {1\over 2}, -{1\over 2})\rightarrow e^{{i\pi\over 4}}( -{1\over 2},{1\over 2}, -{1\over 2})\, ,\nonumber \\
&{~}& ( {1\over 2},- {1\over 2}, -{1\over 2})\rightarrow e^{{i\pi\over 4}}( {1\over 2},- {1\over 2}, -{1\over 2})\, ,\nonumber \\
Q^{\dot\alpha,\dot A}:\quad &{~}& ( -{1\over 2}, -{1\over 2}, -{1\over 2})\rightarrow e^{{3i\pi\over 4}}( -{1\over 2}, -{1\over 2},- {1\over 2})\, ,\nonumber\\
&{~}& ( -{1\over 2}, {1\over 2}, {1\over 2})\rightarrow e^{-{i\pi\over 4}}( -{1\over 2},{1\over 2}, {1\over 2})\, ,\nonumber \\
&{~}& ( {1\over 2},- {1\over 2}, {1\over 2})\rightarrow e^{-{i\pi\over 4}}( {1\over 2},- {1\over 2}, {1\over 2})\, ,\nonumber \\
&{~}& ( {1\over 2}, {1\over 2}, -{1\over 2})\rightarrow e^{{-i\pi\over 4}}( {1\over 2}, {1\over 2}, -{1\over 2})\, .
\end{eqnarray}


Next we will consider the S-duality transformations. The  S-duality is a non-perturbative symmetry of ${\cal N}=4$ supersymmetric SYM theories.
As in \cite{Garcia-Etxebarria:2015wns} we will choose a particular order four  element of the $SL(2,{\mathbb Z})$ S-duality group, which acts on all eight supercharges
$Q^{\alpha A}$ in the same way:
\begin{equation}\label{Scharges}
{\rm S-duality:}\quad Q^{\alpha, A}\rightarrow e^{-{i\pi\over 4}}
Q^{\alpha, A}\, .
\end{equation}
On the remaining eight supercharges $Q^{\dot\alpha ,\dot A}$ S-duality acts in the opposite way:
\begin{equation}
{\rm S-duality:}\quad Q^{\dot\alpha\dot A}\rightarrow e^{{i\pi\over 4}}
Q^{\dot\alpha ,\dot A}\, .
\end{equation}
Combining these transformation rules with the R-symmetry transformations in eq.(\ref{Rcharges}), it immediately follows that by projection on the invariant
charges the combined action of $R\cdot S$, called S-fold projection, leaves
twelve supercharges invariant, i.e. leads to ${\cal N}=3$ supersymmetry.
As we will discuss later, the implementation of the S-duality transformations in the heterotic string will be  quite subtle.




Now let us determine the additional operators, which are invariant under the S-fold projection,
of the non-perturbative ${\cal N}=3$  field theory.
This theory is strongly coupled and there is no Langragian description of this theory\footnote{If $SU(2,2/3)$ is the boundary superconformal algebra,
then the dual bulk theory should be ${\cal N}=6$ $AdS_5$ supergravity which is not the same as ${\cal N}=8$ $AdS_5$ supergravity \cite{Ferrara:1998zt}.
The same projection, which involves an
S-duality transformation leading from ${\cal N}=4$ to 
${\cal N}=3$ supersymmetry on the boundary, reduces in the bulk
${\cal N}=8$  to ${\cal N}=6$ supergravity.}.
The crucial point is that in   super Yang-Mills, the S-fold projection eliminates the massless fields altogether. 
According to     \cite{Ferrara:1998zt},  the singleton of the ${\cal N}=3$ field theory is described by a superfield strength and an $SU(3)$ triplet $V_i(x,\theta)$, which decomposes
under ${\cal N}=1$ into a vector multiplet and three chiral multiplets.
However the singleton of scaling dimension $h=1$ is itself not invariant under the S-fold projection.
The invariant ${\cal N}=3$ operators possess scaling dimension $h=2p$ and have the following form:
\begin{equation}
\Phi^{2p}={\rm Tr}(V_{i\, 1}\bar V_{i\, 2}\dots V_{i\, 2p-1}\bar V_{i\, 2p})\, .
\end{equation}
The lowest operator, denoted as  $W_{{\cal N}=3}\equiv \Phi^2$, has scaling dimension $h=2$  and specifically is given by
\begin{equation}
W_{{\cal N}=3}={\rm Tr}(V_{i}\bar V^j -{1\over 3}\delta_i^jV_k\bar V^k)\, .
\end{equation}
This operator is nothing else than the 
  energy momentum tensor (i.e. supercurrent) of the ${\cal N}=3$ field theory.
 It corresponds to the massive  ${\cal N}=3$ spin-two super-Weyl multiplet
  with, as already shown before,  the following decomposition of massive component fields:
\begin{equation}\label{superWeyl}
W_{{\cal N}=3}:\quad {\rm Spin}(2) + {\underline{6}}\times {\rm Spin} ( 3/2 )  + ({\underline{14}}+
{\underline{1}})\times {\rm Spin}(1) + ({\underline{14'}}+{\underline{6}})\times {\rm Spin} (1/2) + {\underline{14}}\times {\rm Spin} (0)\, .
\end{equation}
It
 contains $n_B=n_F=64 $ degrees of freedom.  
 These massive states were also anticipated  before in \cite{deWit:1978pd}.
 The multiplicities in front of the different fields denote the representations with respect to the group $USp(6)$,
 which is the relevant automorphism group of the massive ${\cal N}=3$  super algebra.
 Note that the ${\cal N}=3$ massive spin-two multiplet is not the same as the ${\cal N}=4$ massive spin-two multiplet.
 
In addition,  one can also consider the  $h=2$ operator
 \begin{equation}
w_{{\cal N}=3}={\rm Tr}V_iV_j\, .
 \end{equation}
 This operator correspond to a massive spin=3/2 supermultiplet of ${\cal N}=3$ with the following field content:
 \begin{equation}
 w_{{\cal N}=3}:\quad {\rm Spin} ( 3/2 ) + {\underline {6}}\times {\rm Spin}(1)+ {\underline{14}}\times {\rm Spin} (1/2)+{\underline{14'}}\times {\rm Spin}(0)\, .
 \end{equation}
 However this operator, which contains $n_B=n_F=32 $ fields, is not invariant under the $R\cdot S$ projection.
 
 The operators $W_{{\cal N}=3}$ and $w_{{\cal N}=3}$ immediately follow from the decomposition of the ${\cal N}=4$ supercurrent $W_{{\cal N}=4}$. 
 As already shown, 
 this current contains $n_B=n_F=128 $ component fields:
 \begin{equation}\label{superWeyln4}
W_{{\cal N}=4}:\quad     {\rm Spin}(2) + {\underline{8}}\times {\rm Spin} ( 3/2 )  + {\underline{27}}
\times {\rm Spin}(1) + {\underline{48}}\times {\rm Spin} (1/2) + {\underline{42}}\times {\rm Spin} (0)        \, ,
\end{equation}
with multiplicities  with respect to the massive ${\cal N}=4$ automorphism group $USp(8)$.
$W_{{\cal N}=4}$ decomposes under the ${\cal N}=3$ currents as\footnote{All these multiplets can be regarded as massless multiplets in the five-dimensional, holographically dual
${\cal N}=8$ and ${\cal N}=6$ supergravity theories.}:
  \begin{equation}
W_{{\cal N}=4}=          W_{{\cal N}=3}+2\times w_{{\cal N}=3}               \, .
\end{equation}
As said before, only $W_{{\cal N}=3}$ survives the S-fold projection.

 \subsection{Non-standard massive W-supergravities as double copy constructions:}

So far we did not consider ${\cal N}=3$ rigid field theory in one of the factors,
say in QFT$({{\cal N}_R})$, of the double copy construction. 
The W-supergravities always have at least one QFT$({\cal N}=3)$ field theory in one of the factors of the double construction:
     \begin{equation}  
    {\rm W-supergravity} \quad=\quad {\rm QFT}({{\cal N}_L})  \otimes   {\rm    QFT}({{\cal N}_R}=3)
                                                                                    \end{equation}
As we discussed in the last section,
 in  QFT$({\cal N}=3)$   the lowest operator (which is invariant under the S-fold twist) is the supercurrent (energy momentum tensor)  of the ${\cal N}=3$  field theory.
  Strictly speaking the supercurrent (being the system superconformal) has the same structure as the super-Weyl
multiplet, which contains the spin-two and the other gauge fields of the superconformal algebra.
Therefore, in the tensor product, the lowest field appearing is 
the ${\cal N}=3$ massive spin-two
 Weyl multiplet  with $n_B=n_F=64$ massive states. 
 Due to the constraint $h_L = h_R$,  the Weyl multiplet of QFT$({{\cal N}_R}=3)$     must be tensored also with  massive 
 fields of the same mass, namely also with the spin-two massive Weyl  supermultiplet from QFT$({{\cal N}_L})$.
   This leads to a massive spin-four field 
   $ \Phi^4_{L+R,{\cal N}_L+3}$ as lowest 
    possible operator in massive W-supergravity:
   \begin{equation}
{\rm W-supergravity}:\quad \Phi^4_{L+R,{\cal N}_L+3}=W_{L, {\cal N}_L} \otimes W_{R,{\cal N}_R=3} \, .
\end{equation}

Knowing this we can now proceed and give an overview over the following five non-standard,  massive W-supergravity theories:

\vskip0.2cm
\noindent
(i) ${\cal N}_L=4$, ${\cal N}_R=3$. This leads to massive ${\cal N}=7$ W-supergravity.
   Due to the level matching constraint, we have to take the massive spin-2 multiplet of ${\cal N}=4$ in the tensor product, which contains $n_B=n_F=128$ massive states.
    So the relevant tensor product leads to a
                   massive spin-4 of ${\cal N}=7$. It contains $n_B=n_F=16384 $ massive states. This theory and its spectrum will be  discussed in more detail in section \S\ref{wn7}.

\vskip0.2cm
\noindent
(ii) ${\cal N}_L={\cal N}_R=3$. This leads to massive ${\cal N}=6$ W-supergravity.

\vskip0.2cm
\noindent
(iii) ${\cal N}_L=2$, ${\cal N}_R=3$. This leads to massive ${\cal N}=5$ W-supergravity.
     
\vskip0.2cm
\noindent
(iv) ${\cal N}_L=1$, ${\cal N}_R=3$. This leads to massive ${\cal N}=4$ W-supergravity.
     
\vskip0.2cm
\noindent
(v) ${\cal N}_L=0$, ${\cal N}_R=3$. This leads to massive ${\cal N}=3$ W-supergravity. It is the only case which can be constructed as a heterotic W-string.
Its heterotic spectrum will be  discussed in more detail in section \S\ref{w3heterotic}.

\vskip0.5cm
\noindent
All double copy constructions, standard massless supergravity theories as well
as massive W-supergra\-vi\-ty theories, have in fact the same interpretation,   we are always tensoring superconformal
field theories, for the spin-one case they are Yang-Mills, for the spin-two case they are conformal (Weyl) gravity.
They are in any case local field theories.  The result is that for spin-one tensoring we have massless spin-two gravity,
while for the spin-two tensoring we have massive  spin-four W-supergravity, which therefore exists for ${\cal N}=7$.

Finally note that the cases 
${\cal N}_L+{\cal N}_R=3,4,5,6$
can be also realized as standard supergravities with massless spin-two.
So for these cases there exist two inequivalent double copies:  standard spin-two supergravity
and massive spin-four  W-supergravity.
Only for ${\cal N}=7$ there exist only the massive spin-four  double copy W-supergravity construction.

   \subsection{Massive ${\cal N}=7$ W-supergravity}\label{wn7}

  In this section we focus on the
     the construction of the  ${\cal N}=7$  W-supergravity theory. 
It  is constructed as a   tensor product of (${\cal N}=4$  ) $\times$ (${\cal N}=3$) field theories. 
Hence  the first, massive  higher spin field in the ${\cal N}=7$ W-supergravity  theory is obtained  from the product
$ W_{L, {\cal N}_L=4} \otimes W_{R,{\cal N}_R=3}\simeq({\rm Spin}= 2)_{{\cal N}=4} \otimes ({\rm  Spin}= 2)_{{\cal N}=3}$. This tensor product starts with one spin-four field and has the following decomposition:
\begin{eqnarray}\label{II=7}
&~& \Bigl({\rm Spin}(2) + 8 \times {\rm Spin}({3\over 2}) +     27 \times {\rm Spin}(1)+    48 \times {\rm Spin}({1\over 2}) + 42 \times {\rm Spin}(0)\Bigr)_{{\cal N}=4} 
   \nonumber\\
  &~&~~~~~~~ 
 \otimes  \Bigl(
 {\rm Spin}(2) + 6\times {\rm Spin} ( 3/2 )  + 15\times {\rm Spin}(1) + 20\times {\rm Spin} (1/2) + 14\times {\rm Spin} (0)
 \Bigr)_{{\cal N}=3} 
\nonumber\\
& = &
   \Bigl({\rm Spin}( 4) + {\underline{14}}_1\times {\rm Spin} ({7\over 2}) + ({\underline{90}}_2+{\underline{1}})\times {\rm Spin}(  3 )+ ({\underline{350}}_3+{\underline{14}}_1) 
   \times {\rm Spin}({5\over 2})
   \nonumber\\
  &~&~~~~~~~ 
  +({\underline{90}}_2+{\underline{910}}_4)\times {\rm Spin}(  2 )
+ ({\underline{350}}_3+{\underline{1638}}_5) \times {\rm Spin}({3\over 2}) + ({\underline{2002}}_6+{\underline{910}}_4)\times{\rm Spin}(1) 
  \nonumber\\
  &~&~~~~~~~ 
+ ({\underline{1430}}_5+{\underline{1638}}_5)\times {\rm Spin} ({1\over 2} )+ {\underline{2002}}_6 \times {\rm Spin}(0)\Bigr)_{{\cal N}=7}\, ,
\end{eqnarray}   
where the two factors contain $2^8$ and $2^7$ states, respectively.
 The multiplicities of these massive states fall into representations of the group $ USp(14)$, which is the automorphism group of ${\cal N}=7$ supersymmetry for massive
      states, where
the 
label means an irreducible, traceless, n-fold $USp(14)$ antisymmetric representation.
It contains $n_B+n_F=2^{15}=32768$  degrees of freedom. Note that this massive multiplet does not constitute a proper ${\cal N }=8$ supermultiplet, but it only transforms under
   ${\cal N }=7$ supersymmetry transformations. 
           It contains   1000 massive graviton-like, spin-two  fields. Since it arises from the tensor product of a strongly coupled ${\cal N}=3$ SYM theory, it is 
      conceivable that this spin-four multiplet also only exists in a strongly coupled theory without a Langrangian description.
      As we will show in the next section, the ${\cal N}=7$ spin-four multiplet arises at the first mass level in the corresponding superstring S-fold construction.
      Furthermore note that the ${\cal N}=7$ W-supergravity theory can be also obtained as S-fold from ${\cal N}=8$ supergravity after including
      an ${\cal N}=8$ spin-four multiplet in a suitable way.

\section{String S-fold construction - Implementation of R-symmetry and S-duality transformations  in heterotic and type II string theory}

Now we want to construct analogous S-folds in type II and heterotic string theories. We call these  theories W-strings, since they do not contain any massless
excitations, but only massive string excitations of higher spin. Their effective description is then supposedly given in terms of the massive W-supergravity theories,
discussed before.

Recall that the S-fold projection in field theory include a discrete R-symmetry transformations times a discrete S-duality transformations.
As we will now see, the field theory R-symmetry transformations will be realized as special kind of T-duality transformations in string theory.
  In fact,  discrete elements of the
T-duality group act   like discrete R-symmetry transformations, which are subgroups of the $SO(6)_R$ automorphism group of the ${\cal N}=4$
 supersymmetry algebra, as it was discussed in
 \cite{Ibanez:1992hc,Ibanez:1992uh}. Hence the R-charges of the space-time fields are closely related to the internal modular T-duality transformation.

In the following we will first consider the right-moving 
 world-sheet degrees of freedom of the heterotic or type II string on $T^6$. The internal coordinates of the six-torus are denoted by $Y^I$, with
$I=1,\dots , 6$. In the four-dimensional uncompactified space-time we go to the light-cone gauge, and the relevant transversal spatial coordinates, corresponding to the transversal
little group 
$SO(2)_T$,
are denoted by $X^i$ with $i=1,2$.
In the 10-dimensional, right-moving sector of the heterotic string we have a world-sheet theory with local $n=1$ world-sheet 
supersymmetry. The right-moving word-sheet fields of the R-NS fermionic string  are  then given as:
\begin{equation}
X^i(z)\, ,\quad Y^I(z)\, ,\quad\psi^i(z)\, ,\quad \lambda^I(z)\, ,\quad \phi(z)\,.
\end{equation}
The $\psi^i(z)$ and the $ \lambda^I(z)$ are  world-fermions and $\phi(z)$ is the bosonized superconformal ghost field.
We can introduce complex coordinates  on the torus, 
\bea
Z^K(z)=Y^{2K-1}(z)+iY^{2K}(z)\, ,\quad K=1,\dots , 3\, .
 \eea 
Furthermore it will be convenient to  introduce complex fermions as 
\begin{equation}
\Psi^0(z)=\psi^1(z)+i\psi^2(z)\, ,\quad\Psi^K(z)=\lambda^{2K-1}(z)+i\lambda^{2K}(z)\, .
\end{equation}
Bosonization of these complex world-sheet fermions is then performed in the standard way:
\begin{equation}
\Psi^0(z)=e^{iH^0(z)}\, ,\quad \Psi^K(z)=e^{iH^K(z)}
\end{equation}
The $H^0(z)$, $H^K(z)$ are four chiral (right-moving) bosons on the world-sheet.

In the 
left-moving sector of the heterotic string there are the following world-sheet fields:
\begin{equation}
\bar X^i(\bar z)\, ,\quad \bar Y^I(\bar z)\, ,\quad \bar Y^{I'}(\bar z)\, .
\end{equation}
Here the $\bar X^i(\bar z)$  the left-moving uncompactified coordinates, the $\bar Y^I(\bar z)$ are the left-moving coordinates on $T^6$ and the $\bar Y^{I'}(\bar z)$ ($I'=1,\dots ,16$)
are the additional bosonic coordinates that are associated to the $U(1)^{16}$ Cartan sub-algebra of the additional heterotic gauge group $G_L$ of rank 16.
In case of the type II sring, the $\bar Y^{I'}(\bar z)$ are absent and instead there are also the left-moving fermions on the world-sheet.

The relevant massless bosonic spectrum  and their associated vertex operators (in the canonical ghost picture) are given as follows:
 \begin{eqnarray}
 {\rm graviton}:&~&G^{( ij)}=\bar\partial \bar X^{( i}(\bar z)\psi^{j)}(z)e^{-\phi(z)} \, ,\nonumber\\
 {\rm anti-sym.~tensor:}&~&B^{\lbrack ij\rbrack}=\bar\partial \bar X^{\lbrack i}(\bar z)\psi^{j\rbrack}(z)e^{-\phi(z)}\, , \nonumber\\
 {\rm dilaton:}&~&\Phi=\bar\partial \bar X^{ i}(\bar z)\psi^{i}(z)e^{-\phi(z)}\,, \nonumber\\
 3~{\rm complex~ structure~moduli~of~T^6:}&~&\tau^K=\bar\partial  Z^{ K}(\bar z)\Psi^{K}(z)e^{-\phi(z)}\, ,\nonumber\\
  3~{\rm Kaehler ~moduli~of~T^6:}&~&\rho^K=\bar\partial \bar Z^{ K}(\bar z)\Psi^{K}(z)e^{-\phi(z)}\, . \label{spectrum}
  \end{eqnarray}

The superpartners of all these states are obtained by performing the operator product of the associated vertex operators with the vertex operators of the four supercharges
of the ${\cal N}=4$ space-time supersymmetry, where the four space-time supercharges $Q^{\alpha,A}$ are coming from the superconformal, right-moving sector of
the heterotic string. 
The spinors  are realized on the world-sheet by the space-time $SO(2)_T$ spin fields 
$S^a(z)$
together
with the internal spin fields $S^A(z)$ of the $SO(6)_R$ R-symmetry group. In the canonical $-1/2$ ghost picture the four supercharges of positive helicity  take the following specific form:
\begin{equation}\label{superchargesp}
\begin{split}
Q^{\alpha= {1\over 2} ,A} &=\oint {dz\over 2i\pi}S^{a={1\over 2}}(z)S^{ A}(z)e^{-{\phi(z)+iH_{\star}(z)\over 2}} \\
	&=\oint {dz\over 2i\pi}e^{{1\over 2}iH^0(z)}e^{iw^{ A}_KH^K(z)}e^{-{\phi(z)+ iH_{\star}(z)\over 2}} \,,
\end{split}
\end{equation}
whereas the four  negative helicity supercharges are given as
\begin{equation}\label{superchargesn}
\begin{split}
Q^{\alpha= -{1\over 2}, A} &=\oint {dz\over 2i\pi}S^{a=-{1\over 2}}(z)S^{ A}(z)e^{-{\phi(z)-iH_{\star}(z)\over 2}} \\
	&=\oint {dz\over 2i\pi}e^{-{1\over 2}iH^0(z)}e^{iw^{ A}_KH^K(z)}e^{-{\phi(z)- iH_{\star}(z)\over 2}} \,.
\end{split}
\end{equation}
Here the four  vectors $w^{A}_K=\vec w^{ A}=(\pm {1\over 2},\pm {1\over 2},\pm {1\over 2})$ (even number of - signs) are the spinorial weight vectors 
of the ${\underline 4}_s$ representation
of the internal R-symmetry group $SO(6)_R$. Furthermore, $\phi(z)$ is the bosonised superghost field and $H_\star(z)$ is the bosonisation of the longitudinal (lightcone) spacetime directions. 
Here the index $a={1\over 2}(-{1\over 2})$ corresponds to (anti)-spinor indices of the transversal little group $SO(2)_T$. Note that the index $a$ is not identical to the spinor index
$\alpha$ (or $\dot\alpha$) of the four-dimensional Lorentz group $SO(1,3)$, which was introduced in the previous section.\footnote{The precise relation between $a$ and $\alpha$
or $\dot
\alpha$ is as follows: decomposing $SO(1,3)$ as $SO(1,3)\supset SO(2)_T\times SO(2)_L$, the two dimensional Lorentz spinors with
spinor indices  $\alpha={1\over 2},-{1\over 2}$ have the following
charges $(a,b)$ under $SO(2)_T\times SO(2)_L$:
\begin{equation}\label{spinora}
\alpha=1/2:\quad (a,b)=({1\over 2},{1\over 2})\, ,\qquad\alpha=-1/2:\quad (a,b)=(-{1\over 2},-{1\over 2})\, ,
\end{equation}
For the two spinors with spinor indices $\dot\alpha$ we get likewise:
\begin{equation}
\dot\alpha=1/2:\quad (a,b)=({1\over 2},-{1\over 2})\, ,\qquad\dot\alpha=-1/2:\quad (a,b)=(-{1\over 2},{1\over 2})\, ,
\end{equation}}
Also note that we have  have adopted in eqs.(\ref{superchargesp})  and eqs.(\ref{superchargesn})
the convention for the supercharges $Q^{\alpha,A}$ to picking an even number of minus signs in $\vec w^{ A}$.
This implies the transversal spinor index $a$ is correlated with the longitudinal spinor index $b$, as it is indicated in eq.(\ref{spinora}).
For completeness we also list the expressions for the eight conjugate supercharges $Q^{\dot\alpha ,\dot A}$:
\begin{equation}\label{superchargespc}
\begin{split}
Q^{\dot\alpha= {1\over 2}, \dot A} &=\oint {dz\over 2i\pi}S^{a={1\over 2}}(z)S^{ \dot A}(z)e^{-{\phi(z)-iH_{\star}(z)\over 2}} \\
	&=\oint {dz\over 2i\pi}e^{{1\over 2}iH^0(z)}e^{iw^{ \dot A}_KH^K(z)}e^{-{\phi(z)- iH_{\star}(z)\over 2}} \,,
\end{split}
\end{equation}
\begin{equation}\label{superchargesnc}
\begin{split}
Q^{\dot \alpha= -{1\over 2}, \dot A} &=\oint {dz\over 2i\pi}S^{a=-{1\over 2}}(z)S^{\dot  A}(z)e^{-{\phi(z)+iH_{\star}(z)\over 2}} \\
	&=\oint {dz\over 2i\pi}e^{-{1\over 2}iH^0(z)}e^{iw^{\dot  A}_KH^K(z)}e^{-{\phi(z)+iH_{\star}(z)\over 2}} \,.
\end{split}
\end{equation}
Finally, the type II supercharges from the left-moving sector are constructed in the same way as their right-moving counterparts.

Now we want to implement the R-symmetry transformations as well as the S-duality transformations of ${\cal N}=4$ field theory, which we have described in the last section,
 in the heterotic and type II string. 
 { To recall, in field theory the S-fold projection was defined by looking at the discrete R- and S-duality charges of the various fields, where the R-charges originate from
 the internal symmetries, and the S-duality charges are determined by the helicity degrees of freedom of the fields. Then a proper linear combination of the R-  and S-charges defines
 the S-fold projection, where one keeps only the states invariant under the combined, discrete transformation. In string theory, the same procedure can be performed:
 all massless and massive states are classified by the same R- and S-charges as in field theory. Therefore one can perform the same truncation of the massless and massive
 spectrum as in field theory. Here we want to propose an interpretation how this truncation can be regarded as certain rotations on the left- and right-moving world-sheet
 degrees of freedom of the heterotic and the type II string theories. }
 There will be two  essential requirements:
 
 \vskip0.2cm
 \noindent (i): The action on the four right-moving supercharges of the heterotic/type II string is like in field theory.
 
 \vskip0.2cm
 \noindent (ii): There is no action on the four left-moving supercharges of the type II string.
 
 \vskip0.2cm
 \noindent Therefore, from the word-sheet point of view, the string construction has to be completely left-right asymmetric, where the ${\mathbb Z}_4$ S-fold projection
 must only act on the right-moving string degrees of freedom, but not at all on the left-moving degrees of freedom. 
 As we will now discuss we will satisfy these two requirements in the following way:
 
 \begin{itemize}
 
 \item
 {\bf R-symmetry:}   Here we can follow the
 construction in \cite{Florakis:2017zep} and also in     \cite{Condeescu:2013yma}.
 The   ${\mathbb Z}_4$ rotation acts on the internal right-moving six coordinates plus the corresponding
 world-sheet fermions, but is leaving the left-moving internal coordinates and the left-moving internal world-sheet fermions invariant. This asymmetric transformation
 just corresponds to a twist by a T-duality transformations, which means that the internal background can be view as a special kind of non-geometric T-fold space.
All massless geometric moduli are projected out by this transformation.

 \item
 {\bf S-duality:} Here { we will propose that} the  asymmetric ${\mathbb Z}_4$ projection will be realized as a  left-right asymmetric ${\mathbb Z}_4$ rotation on the two transversal
 right-moving space-time coordinates plus the corresponding world-sheet fermions. The left-moving space-time coordinates and world-sheet fermions
 are not rotated. 
 { Therefore, although not being compactified, the transversal space coordinates of the closed string are treated in a  left- and right-asymmetric way. 
 This is a new and non-standard feature of the string theory S-fold construction, which means that the closed string boundary conditions with respect to $\sigma$- and 
 $\tau$-translations are chosen with a different ${\mathbb Z}_4$-phase factor for the left- and right-moving coordinates (see  \cite{Lust:2010iy}  for a discussion on this issue).
 The different treatment of the left- and right-moving uncompactified string coordinates can be also motivated in the framework of double field theory.
 Performing the left-right asymmetric projection basically means that also the uncompactified space is a non-geometric background. As we will see, the projection does not allow
 anymore for massless states, and in particular there is non massless graviton anymore after the S-fold projection. Therefore the background is completely rigid.
 Nevertheless all states can be still classified according their helicity quantum numbers and the transversal Lorentz group still acts in the correct way on all
 states. In this sense, Lorentz invariance will be still preserved.
 In other words, the 
 charges of all states under the left-right asymmetric rotations agree with the S-duality (helicity) charges of the corresponding fields in field theory,
 and the projection on the invariant states will be the same as in field theory.}

 More specifically, in the light-cone gauge the asymmetric ${\mathbb Z}_4$ projection includes three simultaneous actions on the physical fields:
 first it will act on the axion-dilaton S-field in a non-trivial way and, as an effect of it, the S-field is projected out
 and the theory is non-perturbative and strongly coupled.  Second, the asymmetric ${\mathbb Z}_4$ will also act on the transversal graviton G-field; we will call 
 this transformation G-transformation, 
 which is very similar to Ehlers transformations in General Relativity.   
 Since the massless graviton is not-invariant under the G-transformation, it will be projected out as well, and
 the theory does not contain any physical graviton state anymore. Hence we call this theory a topological string theory.   
 Note that the combined action on the S-field and on the G-field becomes an element
 of the Geroch group. It is nothing else than a T-duality transformation, when further going down from four to two dimensions, as it was
 discussed in \cite{Florakis:2017zep}. However unlike in \cite{Florakis:2017zep}, we have to augment in 4D string theory the dilaton and metric transformations by a third type of projection,
 that we denote it by H-transformation, and which acts on the helicity degrees of freedom and ensures four-dimensional Lorentz-invariance. This H-transformation
 was not performed in \cite{Florakis:2017zep}, with the result that the two-dimensional theories in \cite{Florakis:2017zep} in general
 posses chiral $({ \cal P},{ \cal Q})$ supersymmetry and are not up-liftable to a four-dimensionional
 Lorentz-invariant theory.
 We will comment on this interplay between four and two dimensions more below. So upshot of this discussion will be that the combined action of S- and G- and H-transformations
 will correspond to the S-duality transformations of supersymmetric field theory, which act on the supercharges in the  way as described in section \S\ref{sfoldft}
 
 \end{itemize}

Before we start the
explicit discussion,  let us also mention that in contrast to the two-dimensional models, for which a full modular invariant partition functions can be constructed \cite{Florakis:2017zep},
 the non-perturbative, four-dimensional string models do not allow for an immediate construction of a one-loop partition function. Furthermore we will also not consider
 twisted sectors, where in two dimensions additional supercharges have emerged in the completely left-right asymmetric orbifold constructions.
 Here will will only discuss the  ${\mathbb Z}_4$ invariant, but untwisted sector of the theory. At the same time we 
 will assume here that possible twisted sectors do not provide any further supercharges in four dimensions.

\subsection{R-symmetry and T-duality twist}

The implementation of the ${\mathbb Z}_4$ R-symmetry rotation is relatively straightforward and we will follow the construction in \cite{Florakis:2017zep}. 
Here we will  give a few more details in terms of the non-geometric interpretation as T-fold.
The full T-duality group for the heterotic string on $T^6$ is given by the discrete group $SO(6,22;{\mathbb Z})$.
For our purpose it is enough 
to work out the action of the various target space duality transformations on the internal string coordinates
of $T^6=\prod_{K=1}^3T^2_i$.
The relevant T-duality subgroup of  $SO(6,22;{\mathbb Z})$ is given by the following modular group:
\begin{equation}
G=O(2,2;{\mathbb Z})^3         =\prod_{K=1}^3\bigl\lbrack SL(2)_{\tau_K}\times SL(2)_{\rho_K}\times \mathbb{Z}_2^{\tau_K\leftrightarrow \rho_K}\times \mathbb{Z}_2^{\tau_K\leftrightarrow -\bar\rho_K}
\bigr\rbrack\, .
\end{equation}
Here the $\rho_K$ and $\tau_K$ are the three K\"ahler respectively complex structure moduli of the three subtori $T^2_K$.

Let us first  restrict the discussion
to one particular  subtorus.
Particular  $O(2,2)$ transformations act as in general left-right asymmetric transformations  on the internal left- and right-moving coordinates 
$Y^i_L(\bar z) $ and $Y^i_R(z)$ ($i=1,2$) of $T^2$:
\begin{equation}
\vec Y_L\rightarrow {\cal M}_L \vec Y_L\, ,\quad \vec Y_R\rightarrow {\cal M}_R \vec Y_R\, .
\end{equation}
In the following we are interested in those ${\cal M}_{SO(2,2)}$ transformations, for which the  group elements $({\cal M}_L,{\cal M}_R)$
correspond to  discrete rotations 
 $\mathbb{Z}_N^L\times \mathbb{Z}_M^R$ on the torus coordinates.
This  is possible, provided that a faithful embedding of $\mathbb{Z}_N^L\times \mathbb{Z}_M^R$
into $O(2,2)$ can be found (see \cite{Condeescu:2013yma} for more details).
Introducing a complex coordinate  on the torus, 
\bea
Z(z,\bar z)=Y^1(z,\bar z)+iY^{2}(z,\bar z)=Z_L( \bar z)+Z_R(z)\, ,
 \eea 
 the $O(2,2)$ transformations then act  as particular discrete rotations on the complexified left- and right-moving coordinates:
\bea
Z_L( \bar z)  \rightarrow e^{2i\pi/N_L} Z_L( \bar z)\, ,\qquad Z_R(z)\rightarrow  e^{2i\pi/N_R} Z_R(  z)\, .
 \eea

Next consider the corresponding right-moving, complex world sheet fermions
$\Psi^{K}$.
Their transformation behavior can be deduced from
the requirement  that the right-moving world sheet supersymmetry
commutes with the target space modular transformations.
This requirement follows from the fact that the action of the
right-moving supercurrent connects equivalent (picture-changed)
physical string states.
The right-moving (internal)
world sheet supercurrent has the form
\begin{equation}T_F(z)=\sum_{K=1}^3(\Psi^{K}_R\partial \bar Z^{K}_R+\bar \Psi^{K}_R\partial
Z^{K}_R)(z).
\end{equation}
Demanding the supercurrent to be invariant
under modular transformations, one derives that  the complex fermions $\Psi(z)$ of each subtorus transform only under the rotations of the right-movers as
\begin{equation}
\Psi(z)\rightarrow e^{2i\pi/N_R}
\Psi(z).
\end{equation}
It follows that these transformations
act on the three two-dimensional bosons $H$ as
\begin{equation}
H(z)\rightarrow H(z)+2\pi/N_R\end{equation}

\vskip0.8cm

\noindent
One can consider three particular cases: (i) $N_L=N_R\neq1$, (ii) $N_L=-N_R\neq1$ and (iii) $N_L=1,N_R\neq1$.
Case (i), the symmetric rotation, corresponds to a transformation, which only acts on the complex structure modulus $\tau$ but not on the K\"ahler modulus
$\rho$. Conversely the asymmetric case (ii) can be realized by T-duality transformations that act only on $\rho$ but not on $\tau$.
Finally, the completely asymmetric case (iii), for which the left-moving degrees of freedom are inert, corresponds to transformations which simultaneously act on
$\rho$ and on $\tau$. For the reasons explained above, this is the case of interest for us, and to be specific we choose $-N_L=N_R=4$. 
In order to realize this asymmetric projection one has  to  freeze the moduli to their self-dual values, i.e. $\tau=\rho=i$.
It follows that the corresponding moduli fields are projected out of the physical spectrum of the string.

Now we are ready to examine the modular transformation properties
of the  space-time supercharges $Q^{\alpha, A}$ in 
eqs.(\ref{superchargesp}) and (\ref{superchargesn}). Namely they transform with a particular common phase under these T-duality transformations. Specifically we obtain 
 that 
 the  supercharges in the right-moving sector of the heterotic string transform in the same way as:
\begin{equation}
 T-{\rm duality}:\quad Q^{\alpha, A}
\rightarrow e^{-{i\pi(w_1^A+w_2^A+w_3^A)\over 2}}
 Q^{\alpha,A}\,. \label{ttrans}
\end{equation}

 These transformations
 agree with the  R-transformations in field theory, described above (see eq.(\ref{Rcharges})).  In fact, the discrete ${\mathbb Z}_4$ elements of the
T-duality group act   like discrete R-symmetry transformations, which are subgroups of the $SO(6)$ automorphism group of the ${\cal N}=4$
 supersymmetry algebra. Hence the charges of the space-time fields under the target space modular transformation  are closely related to their corresponding R-charges,
 as it was already discussed in
 \cite{Ibanez:1992hc,Ibanez:1992uh}

\subsection{S-duality twist}

  \vskip0.4cm
\noindent{\bf S-duality:}

\vskip0.2cm
\noindent
S-duality  \cite{Font:1990gx} exists in the four-dimensional heterotic string and also in the NS sector of the type II strings as non-perturbative symmetry that acts
on the four-dimensional dilaton-axion field, denoted by $S=a+ie^{-2\Phi}$, as $SL(2,{\mathbb Z})_S$ transformations ($\Phi$ is the dilaton field and $a$ is the dualized
four-dimensional $B_{\mu\nu}$ field):
 \begin{equation}\label{sdualstring}
 S\rightarrow{ a\,S+ b\over  c\,S+ d}\, ,\quad a d- b c=1\, .
 \end{equation}
In the four-dimensional effective field theory, 
 the non-perturbative S-duality transformations  map  the perturbative heterotic states with electric charges onto the non-perturbative states with magnetic charges and vice versa.
S-duality also maps the electromagnetic field strengths and the dual field strength into each other. From a ten-dimensional string point of view, S-duality exchanges the elementary 
strings with the solitonic five-branes. During the compactification process on $T^6$ one is wrapping the five-branes around the internal 5-cycles, and then the electric-magnetic 
S-duality
in four dimensions becomes manifest.

    \vskip0.4cm
\noindent{\bf World-sheet action of S-duality:}

\vskip0.2cm
\noindent

   In addition to its non-perturbative action, S-duality also acts on the world-sheet fields in a particular, non-trivial way, which will be important for the construction
 of the S-folds. For that purpose let us introduce also a complex coordinate for the uncompactified, transversal coordinates $X^1$ and $X^2$:
\bea
Z^0_L( \bar z)=X^1(\bar z)_L+iX^2(\bar z)_L\,,\quad Z^0_R( z)=X^1(z)_R+iX^2(z)_R\, .
 \eea 
 Then the complex S-field is associated to the following vertex operator:
 \begin{equation}
 {\rm S-field:}~~S=\bar\partial \bar Z^0_L( \bar z)\Psi^{0}(z)e^{-\phi(z)}\, .
 \end{equation}
 As already observed in  \cite{Font:1990gx}, this operator  is in complete analogy  to the marginal operator of the internal K\"ahler modulus $\rho$ of an internal 2-torus.
 Indeed, from the world-sheet conformal field theory point of view, the modes $\rho$ and $S$ are very similar, both corresponding to
 $\bar \partial Z_L\partial Z_R$ marginal deformations with respect to the compactified  or uncompactified spatial coordinates (with periodic boundary conditions in the internal
 directions). 
 The
  action of the S-duality transformations  
 have the same left-right asymmetric  effect on the uncompactified transversal coordinates  $Z^0_L$ and $Z^0_R$ and world-sheet fermions as the  modular T-duality transformations on the 
 K\"ahler modulus $\rho$ have on the internal
 world-sheet fields. 
 In fact if we further compactified the two transverse coordinates $Z^0$ and $\bar Z^0$ on a torus, the parallelism between $\rho$ and  $S$ would be strict, but of course, that
 is now not the case.\footnote{So compared to F-theory, where S-duality originates from the modular transformations on the complex structure
 of the auxiliary type IIB torus, the heterotic S-duality can be seen to have its origin from the ''torus'' that describe the transversal spatial coordinates in four dimensions.} 
 The fact, that S-duality is  reduced to T-duality when putting the theory on a 2-torus was also discussed  in the context of four-dimensional ${\cal N}=4$ super Yang-Mills
 theory  \cite{Harvey:1995tg}.

    \vskip0.4cm
\noindent{\bf World-sheet action of G-duality:}

\vskip0.2cm
\noindent
 As already advertised,  in addition to S-duality we introduce a another kind of duality transformation, which we call G-duality. It acts on the two light-cone degrees of freedom of the graviton field in uncompactified
 four-dimensional Minkowski space. 
  G-duality transformations act on the world-sheet fields $Z^0_L$ and $Z^0_R$ in a left-right-symmetry way, in complete analogy to the action of $SL(2)_\tau$ duality transformations on 
the  internal
 world-sheet fields of a 2-torus.
 Therefore, in analogy to  the complex structure modulus $\tau$  on the compact two-torus, let us combine the two transversal degrees of freedom of the four-dimensional graviton field  into one
 complex graviton field $G$ as follows:
 \begin{equation}
 G={\frac{g_{12}}{g_{11}}}+i\ {\frac{\sqrt{\det g}}{g_{11}}}
 \, .
 \end{equation}
 Now $SL(2)_g$ acts on this field in the standard way:
 \begin{equation}\label{vdual}
 G\rightarrow{ a\,G+ b\over  c\,G+ d}\, ,\quad a d- b c=1\, .
 \end{equation}
 Seen from the point of general relativity, the G-duality transformations  are just a certain kind of large diffeomorphisms. Namely they act like finite,
 four-dimensional   coordinate 
 transformations.
 From the world-sheet point of view
they have the same left-right symmetric action on the uncompactified transversal coordinates and world-sheet fermions as the corresponding modular T-duality transformations on the complex  modulus $\tau$ have on the internal
 world-sheet fields of a compact two-torus.

    \vskip0.4cm
\noindent{\bf Combined world-sheet action of S-duality and G-duality:}

\vskip0.2cm
\noindent

  Like on the two-torus, it also becomes evident that  $SL(2)_S$ and $SL(2)_G$ combine into the group $O(2,2)_{S,G}\simeq SL(2)_S\times SL(2)_G$. 
  This group is a symmetry of the general relativity  and also of the two-dimensional string $\sigma$-model, in case the four-dimensional background possesses two Killing 
  symmetries. The group  $O(2,2)_{S,G}$ is the socalled Geroch group \cite{Geroch:1972yt}  and the string Geroch group was already discussed in 
  \cite{Bakas:1994np,Duff:1994zt} some time ago.
   Moreover,  it follows that 
 group elements of 
  the rank seven group $O(6,6) \times SL(2)_S$ are promoted  by the inclusion of $SL(2)_G$ to certain group elements of the rank eight group $O(8,8)$. This group is identical to the T-duality group of $T^8$, when we compactify to  two dimensions.   
 In general S- and G-duality transformations act the uncompactified, transversal world-sheet fields as
  \bea
Z^0_L( \bar z)  \rightarrow e^{2i\pi/M_L} Z^0_L( \bar z)\, ,\qquad Z_R^0(z)\rightarrow  e^{2i\pi/M_R} Z_R^0(  z)\, ,
 \eea 
and 
 \begin{equation}
\Psi^0(z)\rightarrow e^{2i\pi/M_R}
\Psi^0(z).
\end{equation}
 We can now consider again the three different cases, depending how S-duality and G-duality transformations act on the left- and right-moving complex coordinate $Z^0_L$ and $Z^0_R$.
 (i) $M_L=M_R\neq1$, (ii) $M_L=-M_R\neq1$ and (iii) $M_L=1,M_R\neq1$.
Case (i), the symmetric rotation, corresponds to a transformation, which only acts on the metric $G$ but not on the S-field.
Conversely the asymmetric case (ii) can be realized by S-duality transformations that act only on $S$ but not on $G$.
Finally, the completely asymmetric case (iii), for which the left-moving degrees of freedom are inert, corresponds to transformations which simultaneously act on
$S$ and on $G$. Therefore we call it Geroch twist. Again like for the T-duality transformations, this is the case of interest for us, and to be specific we choose $-M_L=M_R=4$. 
In order to realize this asymmetric projection one has  to  freeze the fields to their self-dual values, i.e. $S=G=i$.
It follows that the massless graviton as well as the S-field is projected out of the physical spectrum

It now  follows that the eight heterotic space-time supercharges in eq.(\ref{superchargesp})  and in eq.(\ref{superchargesn})
transform, depending on their helicity, with an opposite phase factor under S-G-duality transformations, namely
\begin{equation}
 {\rm Geroch~twist}\equiv(S\otimes G)-{\rm duality}:\quad Q^{\alpha={1\over 2} ,A}\rightarrow
e^{ {i\pi\over 4}}Q^{\alpha={1\over 2}, A}\,.\label{stransp}
\end{equation}
and
\begin{equation}
  {\rm Geroch~twist}\equiv (S\otimes G)-{\rm duality}:\quad Q^{\alpha=-{1\over 2} ,A}\rightarrow
e^{- {i\pi\over 4}}Q^{\alpha=-{1\over 2} ,A}\,.\label{stransp1}
\end{equation}
This is almost agrees with the action of S-duality in field theory. However we realize that this transformation acts on the supercharges in a non Lorentz-invariant way.
Namely comparing this transformation with the transformation of the supercharges in field theory, as given in eq.(\ref{Scharges}),
 we see that they do not agree with each other: whereas in field theory, the S-duality phase factor does not depend on the spinor $\alpha$,
the string construction so far leads to opposite phases for the two spinor weights $\alpha=\pm{1\over 2}$. 
Combining it with T-duality
the 
number of conserved supercharges is only six instead of twelve, as required by four-dimensional Lorentz invariance.

   \vskip0.4cm
\noindent{\bf The H-twist:}

\vskip0.2cm
\noindent

As already announced, in order to ensure Lorentz-invariance of the S-duality transformations, we will have to augment the S-G-duality transformations, by a further
twist operation, denoted by $H$, which acts differently on two two components of the fermionic spinor degrees of freedom and  reverses
the phase factor of the positive helicity spinors:
\begin{equation}
H-{\rm twist}:\quad Q^{\alpha={1\over 2} ,A}\rightarrow
e^{- {i\pi\over 2}}Q^{\alpha={1\over 2} ,A}\,,
\quad Q^{\alpha=-{1\over 2} ,A}\rightarrow
Q^{\alpha=-{1\over 2} ,A}\, .
\label{htransp}
\end{equation}

The S-G-H twist is an automorphism  in  string theory and hence it is an allowed projection. It acts as left-right-asymmetric rotation on the transversal coordinate. 
{The charges of all fields under the S-G-H twist are same same as the charges under the discrete S-duality transformation in field theory.}
Therefore the upshot of this discussion is that what we have considered as S-duality in ${\cal N}=4$ supersymmetric field theories is represented in string theory as the product of $S$ and
$H$ transformations. If one in addition requires  to act only the right-moving world-sheet fields, one also has to include the G-transformations:
\begin{equation}
{\rm S}-{\rm duality ~in ~{\cal N}=4~field ~theory} ~~\equiv~~ S\otimes G\otimes H~~{\rm in ~string ~theory}
\end{equation}
Note that the H-twist was not performed in   \cite{Florakis:2017zep}. There only the S-G rotations on the left- and right moving coordinates were performed, which become identical to T-duality
transformations when going to two dimensions. Without only the S-G rotations being performed, the S-fold breaks 4D-Lorentz invariance. However in two-dimensions this is not a problem,
and it results in a chiral 2D theories with $({\cal P},{\cal Q})$ supersymmetry.

\section{Massive spectrum of W-superstrings}
\vskip0.5cm

Now we finally proceed to examine the spectrum of some string theory S-folds. { Their spectra are given by those states which are invariant under the combined discrete 
T-S-G-H rotations.}
The graviton $G^{(ij)}$, the antisymmetric tensor $B^{\lbrack ij\rbrack}$ and the dilaton
$\Phi$ are invariant under the T-duality transformations. Other fields may transform with a particular phase, since the contain either internal world-sheet bosons or
internal world sheet fermions in the vertex operators.
On the other hand the vertex operators of the graviton $G^{(ij)}$, the antisymmetric tensor $B^{\lbrack ij\rbrack}$ and the dilaton
$\Phi$ are not invariant under the S-G-H-transformations. Only fields with only internal degrees of freedom are invariant under S-duality transformations.
Hence when  building the heterotic or type II S-folds with ${\cal N}=3$ or ${\cal N}=7$ supersymmetry, the entire massless perturbative spin-two supermultiplet is projected out of the
string spectrum. Only massive string excitations can survive the S-fold projection. Therefore it is very plausible
 that the  non-perturbative ${\cal N}=7$  supergravity, which arises as effective theory from the string theory, is the  massive, higher spin W-supergravity theory, which we have 
 discussed before.

In the four-dimensional effective field theory, 
 non-perturbative S-duality transformations  map  the perturbative heterotic states with electric charges onto the non-perturbative states with magnetic charges and vice versa.
After the ${\mathbb Z}_4$ projection, the S-field is frozen to its self-dual value $S=i$ and hence the string coupling constant is fixed at strong coupling. It follows that in the S-fold,
the non-perturbative  electric and magnetic states have equal masses, which are comparable with the masses of the perturbative massive string states.
In the following we will nevertheless neglect the non-perturbative states in the discussion of the spectrum of the S-folds, but we will only derive the spectrum of the massive
excited, perturbative string states of the four-dimensional heterotic and type II S-folds that survive the ${\mathbb Z}_4$ projection.
 We will restrict the discussion on the states at the first massive string level.
In particular we will  construct the massive spectrum of ${\cal N}=3$  and  ${\cal N}=7$ type II W-strings.

\subsection{Massive states of the unprojected four-dimensional fermionic string with ${\cal N}=4$ supersymmetry}
\label{spectrferma}

We will start by recalling the massive spectrum of the (left or right-moving) fermionic string after the GSO projection, but before projecting  on the $\mathbb Z_4$ invariant states.
These states will be needed later on, when discussing the ${\cal N}=3,7$  S-folds.

\vskip0.2cm
\noindent{\underline{Bosons:}}
\vskip0.2cm

Splitting the indices into uncompactified and internal indices, 
one  obtains at the first massive level the following massive states (see for example \cite{Blumenhagen:2013fgp}):
\begin{eqnarray}\label{bosons2}
 &~&b^i_{-1/2}b^j_{-1/2}b^I_{-1/2}|0\rangle\, ,\qquad b^i_{-1/2}b^I_{-1/2}b^J_{-1/2}|0\rangle\, ,\qquad b^I_{-1/2}b^J_{-1/2}b^K_{-1/2}|0\rangle\, ,\nonumber\\
&~&b^i_{-3/2}|0\rangle\, ,\qquad   b^I_{-3/2}|0\rangle\,   \nonumber\\
&~& \alpha_{-1}^ib^j_{-1/2}|0\rangle\, ,\quad \alpha_{-1}^ib^I_{-1/2}|0\rangle\, ,\quad\alpha_{-1}^Ib^i_{-1/2}|0\rangle\, ,\quad\alpha_{-1}^Ib^J_{-1/2}|0\rangle\, .
\end{eqnarray}
(Here the $b$'s and the $\alpha$'s are the oscillators of the world-sheet fermions and bosons.)
Collecting all states and putting them into proper massive representations of the four-dimensional little group $SO(3)$ as well as in 
proper representations of the ${\cal N}=4$   $SU(4)$ R-symmetry, one obtains the following massive spectrum:
\begin{equation}\label{N4bos}
{\underline 1}\times {\rm Spin}(2)+({\underline 6}+{\underline 6}+{\underline {15}})\times {\rm Spin}(1)+(2\times\underline1+{\underline {10}}+{\bar{\underline {10}}}+{\underline {20'}})\times {\rm Spin}(0)\, .
\end{equation}
As discussed before, for massive states in ${\cal N}=4$ supersymmetry the R-symmetry group is enhanced from $U(4)$ to $USp(8)\supset U(4)$ with the following branching rules:
\begin{eqnarray}
{\underline 8}&=&{\underline 4}+{\bar{\underline 4}}\, ,\nonumber\\
{\underline {27}}&=&{\underline 6}+{\underline {6}}+{\underline{15}}\, ,\nonumber\\
{\underline {36}}&=&{\underline 1}+{\underline {10}}+\bar{\underline {10}}+{\underline{15}}\, ,\nonumber\\
{\underline{42}}&=&2\times\underline1+{\underline {10}}+{\bar{\underline {10}}}+{\underline {20'}}\, ,\nonumber\\
{\underline{48}}&=&{\underline 4}+\bar{\underline 4}+{\underline {20}}+{\bar{\underline {20}}}
\end{eqnarray}
Then the massive bosons transform under $USp(8)$ as
\begin{equation}\label{N4bossp}
{\underline 1}\times {\rm Spin}(2)+({\underline {27}})\times {\rm Spin}(1)+({\underline{42}})\times {\rm Spin}(0)\, .
\end{equation}

\vskip0.2cm
\noindent{\underline{Fermions:}}
\vskip0.2cm

In ten dimensions, the 128 massive fermions are given by the following string states:
\begin{eqnarray}\label{fermions1}
 (8)_c+(56)_c:\quad b^A_{-1}|a\rangle\, ,\qquad        (8)_s+(56)_s:\quad \alpha^A_{-1}|\dot a\rangle\,   \, .
\end{eqnarray}
In terms of four-dimensional massive spinors this leads to:
\begin{equation}
({\underline 4}+\bar{\underline 4})\times{\rm Spin}(3/2)+({\underline 4}+\bar{\underline 4}+{\underline {20}}+{\bar{\underline {20}}})\times {\rm Spin}(1/2)\, ,
\end{equation}
where in this decomposition each spin 3/2 Rarita Schwinger field in four dimensions contains 4 degrees of freedom and each  spin 1/2 Dirac fermion possess 2 degrees of freedom.
Under $USp(8)$ the massive fermions transform as
\begin{equation}\label{fermions2}
({\underline 8})\times{\rm Spin}(3/2)+({\underline {48}})\times {\rm Spin}(1/2)\, ,
\end{equation}

The bosons in eq.(\ref{N4bos}) together with the fermions in eq.(\ref{fermions2}) build one long, massive ${\cal N}=4$ spin 2 supermultiplet.
It precisely agrees with the super Weyl multiplet $W^2_{{\cal N}=4}$, which is displayed in eq.(\ref{superWeyln4}).

\subsection{Massive states of the ${\mathbb Z}_4$-projected four-dimensional fermionic string with ${\cal N}=3$ supersymmetry}
\label{spectrferm}

Now we perform the   ${\mathbb Z}_4$ S-fold projection  on the above massive spectrum of the right-moving sector of the fermionic string.
Then supersymmetry is broken from   ${\cal N}=4$ to ${\cal N}=3$, the R-symmetry is broken from $SU(4)$ to $U(3)$ and
the supercharge transforms as $\underline{3}$ under the R-symmetry.
We emphasize again that the 
H-projection must be taken into account in order to get the correct Lorenz-invariant spectrum.

\vskip0.2cm
\noindent{\underline{Bosons:}}
\vskip0.2cm

   The 64 invariant   states can be grouped into the   massive representations of the four-dimensional little group $SO(3)$ and of the $SU(3)$ R-symmetry.
  But for the massive states in ${\cal N}=3$ supersymmetry, the R-symmetry group is enhanced from $U(3)$ to $USp(6)\supset U(3)$ with the following branching rules:
\begin{eqnarray}
{\underline 6}&=&{\underline 3}+{\bar{\underline 3}}\, ,\nonumber\\
{\underline {14}}&=&{\underline 3}+\bar{\underline {3}}+{\underline{8}}\, .
\end{eqnarray}
Then the massive bosons transform under $USp(6)$ as
\begin{equation}\label{N3bossp}
{\rm Spin}(2)  + ({\underline {14}}+{\underline {1}})\times {\rm Spin}(1) + {\underline {14}}\times {\rm Spin} (0)
\, .
\end{equation}


 \vskip0.2cm
\noindent{\underline{Fermions:}}
\vskip0.2cm
 Acting with  the ${\mathbb Z}_4$ transformation  on the spinor fields, the invariant states 
build the following 64 massive fermions in four dimensions 
\begin{equation}\label{fermions3}
{\underline {6}}\times {\rm Spin} ( 3/2 ) + ({\underline {14'}}+{\underline {6}})\times {\rm Spin} (1/2) 
\, .
\end{equation}

Note that the bosons in eq.(\ref{N3bossp}) together with the fermions in eq.(\ref{fermions3}) build one  massive ${\cal N}=3$ spin-two supermultiplet, which perfectly agrees
with the spin-two super Weyl multiplet $W_{{\cal N}=3}$, as given in eq.({\ref{superWeyl}).


\subsection{Massive states of the bosonic string}
\label{spectrbos}

For completeness we also need the massive states of the left-moving bosonic string, when we build the heterotic string. Leaving out the internal gauge coordinates, we get 
the following 44 massive states in ten dimensions:
\begin{equation}
(8)_v:\quad \alpha^A_{-2}|0\rangle\, ,\qquad         (35)_0+1_0:\quad \alpha^A_{-1}\alpha^B_{-1}|0\rangle\, .
\end{equation}
They correspond  to the following massive bosonic states in four dimensions:
\begin{equation}\label{bosonicmassive}
{\rm Spin}(2)+6\times {\rm Spin}(1)+21\times {\rm Spin}(0)\, .
\end{equation}
The massive spin-two field is universal and will be present in any background. The 6 massive vectors and the 21 massive
scalars can be regarded as additional matter fields, which appear
in case the internal left-moving is sector is given by a six-dimensional torus.

\subsection{${\cal N}=3$ heterotic W-superstring}\label{w3heterotic}
\vskip0.5cm

This non-perturbative heterotic  theory is given by  the tensor product of the right-moving  massive spectrum of section \S\ref{spectrferm}  times the left-moving spectrum of section 
\S\ref{spectrbos}.
Seen as field theory double construction, it corresponds to the ${\cal N}=3$, W-supergravity, which is a tensor product ${\rm QFT}({\cal N}=3)\otimes {\rm QFT}({\cal N}=0)$.
The universal sector of the ${\cal N}=3$ W-supergravity is obtained by building the tensor product of the left-moving massive spin-two field in eq.(\ref{bosonicmassive})
with the right-moving ${\cal N}=3$ Weyl multiplet:
The massive spectrum has the following form:
\begin{eqnarray}
&~&\hskip-0.8cm{\rm B:}\quad \lbrack{\rm Spin}(2)  + 15\times {\rm Spin}(1) + 14\times {\rm Spin} (0)\rbrack_R     \times\lbrack   {\rm Spin}(2)\rbrack_L   \, ,\nonumber\\
&~&\hskip-0.8cm{\rm F:}\quad \lbrack 6\times{\rm Spin}(3/2)+20\times {\rm Spin}(1/2)\rbrack_R\times\lbrack {\rm Spin}(2)\rbrack_L\, .
\end{eqnarray}
Explicitly performing the tensor product this leads to the following spectrum of massive fermions and bosons:
\begin{eqnarray}
&~&\hskip-0.8cm{\rm B:}\quad \lbrack {\rm Spin}(4)+16\times {\rm Spin}(3)+30\times {\rm Spin}(2)+16\times {\rm Spin}(1)+ {\rm Spin}(0)\rbrack  \nonumber\\
&~&\hskip-0.8cm{\rm F:}\quad \lbrack 6\times{\rm Spin}(7/2)+26\times {\rm Spin}(5/2)+26\times{\rm Spin}(3/2)+6\times {\rm Spin}(1/2)\rbrack \, .
\end{eqnarray}
These states  build two long, massive ${\cal N}=3$  supermultiplets, namely
one ($j=5/2$) with top spin-four  ($n_B+n_F=6\times 64=384$) plus another one ($j=3/2$) with top spin-three ($n_B+n_F=4\times 64=256$), 
where the multiplicities of these states fall into representations of the group $USp(6)\supset U(3)$.
It contains in total $n_B+n_F=640$ degrees of freedom.

If one also includes the additional six spin-one and 21 spin-zero fields of the left-moving bosonic string in eq.(\ref{bosonicmassive}) into account, 
the massive spectrum has the following form:
\begin{eqnarray}
&~&\hskip-0.8cm{\rm B:}\quad \lbrack{\rm Spin}(2)  + 15\times {\rm Spin}(1) + 14\times {\rm Spin} (0)\rbrack_R     \times\lbrack   {\rm Spin}(2)+6\times {\rm Spin}(1)+21\times {\rm Spin}(0)\rbrack_L   \, ,\nonumber\\
&~&\hskip-0.8cm{\rm F:}\quad \lbrack 6\times{\rm Spin}(3/2)+20\times {\rm Spin}(1/2)\rbrack_R\times\lbrack {\rm Spin}(2)+6\times {\rm Spin}(1)+21\times {\rm Spin}(0)\rbrack_L\, .\end{eqnarray}
Then the  tensor product takes following form:
\begin{eqnarray}
&~&\hskip-0.8cm{\rm B:}\quad \lbrack {\rm Spin}(4)+22\times {\rm Spin}(3)+147\times {\rm Spin}(2)+511\times {\rm Spin}(1)+385\times {\rm Spin}(0)\rbrack  \nonumber\\
&~&\hskip-0.8cm{\rm F:}\quad \lbrack 6\times{\rm Spin}(7/2)+62\times {\rm Spin}(5/2)+308\times{\rm Spin}(3/2)+582\times {\rm Spin}(1/2)\rbrack \, .
\end{eqnarray}
These states  build a reducible  massive ${\cal N}=3$ spin-four supermultiplet, and 
it contains in total $n_B=n_F=2816$ degrees of freedom.

\subsection{${\cal N}=7$ type II W-superstring}

\vskip0.5cm

The ${\cal N}=7=4_L+3_R$  non-perturbative type II W-superstring is given as the product of a left-moving times a right-moving fermionic string with four respectively three space-time
supercharges.
It is  strongly coupled and it does not contain any massless fields. The massive sector 
is given by  the tensor product of the right-moving massive spectrum of section \S\ref{spectrferm}   times the left-moving spectrum of section \S\ref{spectrferma} in the following way: 
\begin{eqnarray}
&~&\hskip-0.8cm{\rm B:}\quad \lbrack {\rm Spin}(2)+15\times {\rm Spin}(1)+14\times {\rm Spin}(0)\rbrack _R    \times\lbrack {\rm Spin}(2)+27\times {\rm Spin}(1)+42\times {\rm Spin}(0)\rbrack_L
 \, ,\nonumber\\
 &~&\hskip-0.8cm\quad \lbrack 6\times{\rm Spin}(3/2)+20\times {\rm Spin}(1/2)\rbrack_R\times\lbrack
 8\times{\rm Spin}(3/2)+48\times {\rm Spin}(1/2)
 \rbrack_L\nonumber\\
&~&\hskip-0.8cm{\rm F:}\quad \lbrack 6\times{\rm Spin}(3/2)+20\times {\rm Spin}(1/2)\rbrack_R\times\lbrack 
{\rm Spin}(2)+27\times {\rm Spin}(1)+42\times {\rm Spin}(0)
\rbrack_L\, ,\nonumber\\
&~&\hskip-0.8cm\quad \lbrack{\rm Spin}(2)+15\times {\rm Spin}(1)+14\times {\rm Spin}(0)\rbrack _R    \times\lbrack 
 8\times{\rm Spin}(3/2)+48\times {\rm Spin}(1/2)
\rbrack_L
 \, .
\end{eqnarray}
Altogether we explicitly derive in this way the  following ${\cal N}=7$ spectrum of massive fermions and bosons:
\begin{eqnarray}
&~&\hskip-0.8cm{\rm B:}\quad \lbrack
{\rm Spin}( 4)  + 91\times {\rm Spin}(  3 )+1000\times {\rm Spin}(  2 )
 + 2912\times{\rm Spin}(1) + 2002 \times {\rm Spin}(0)
\rbrack \, , \nonumber\\
&~&\hskip-0.8cm{\rm F:}\quad \lbrack 
14\times {\rm Spin} ({7\over 2}) + 364 \times {\rm Spin}({5\over 2})
  + 1988 \times {\rm Spin}({3\over 2})  + 3068\times {\rm Spin} ({1\over 2} )
\rbrack \, .
\end{eqnarray}
These states  build exactly one  long, massive ${\cal N}=7$ spin-four supermultiplet.
It perfectly agrees with the massive, spin-four multiplet in eq.(\ref{II=7}), which we have obtained via the double copy construction of ${\cal N}=7$ W-supergravity and
where the multiplicities of these states fall into representations of the group $USp(14)\supset U(7)$.
The agreement of the string theory construction and the field theory double copy construction provides some convincing evidence that ${\cal N}=7$ W-supergravity indeed
exists as physical theory.
Since the theory does not contain any massless fields, and hence in particular also no massless gravitons and also no massless dilaton, it should be strongly coupled and also be
a kind of topological theory, since there are no possible fluctuations around a given gravitational background.



\section{Conclusions and Outlook}
In this paper we have provided evidence for the existence of new W-supergravity and W-superstring theories. They are built as 
double copy constructions involving the non-perturbative ${\cal N}=3$ supersymmetric Yang-Mills theories, or in the string context, 
containing a right-moving fermionic string with ${\cal N}=3$ space-time supersymmetry. 
In particular we have obtained in this way a massive,
spin-four W-supergravity/superstring with ${\cal N}=7$ (28 supercharges) supersymmetry in four space-time dimensions. This theory
 can be regarded as a non-perturbative S-fold of ${\cal N}=8$ supergravity or, respectively, of the ${\cal N}=8$ type II  superstring theory.
 
 The proposed ${\cal N}=7$ string construction possesses the following three distinct features:
 
 \begin{itemize}
 
 \item The internal space is a non-geometric background, obtained by modding out with the T-duality group. This eliminates all internal massless moduli.
 
 \item { The discrete S-duality transformation corresponds to left-right asymmetric rotation of the uncompactified transversal coordinates. Therefore also the uncompactified space
 can be regarded as non-geometric background.}
  Then, along the uncompactified directions, the S-duality is modded out. Therefore the theory is strongly coupled without a massless dilaton.

 \item 
 In four dimensions there is an additional twist  by large diffeomorphisms. Therefore there is no massless graviton multiplet, and the theory is a massive,
 topological theory. Together with the S-duality, this twist corresponds to a discrete element of the Geroch group.
 Note that the Geroch group  also contains a kind of mirror symmetry in the uncompactified transversal directions, namely the exchange of the S-field with the (transversal)
 graviton field.
  \end{itemize}
 { As a caveat we like to mention that we have discarded possible twisted sectors in our string construction, we have just performed a truncation of the entire string spectrum
 on the states, which are invariant under the discrete asymmetric rotations. Due to their non-perturbative character, we also could not construct a modular
 invariant partition function for the W-superstrings. 
 Therefore there is at the moment no proof that the W-superstrings are fully consistent, full fledged string theories.}

 We also like to emphasize that, although being massive theories, W-supergravities are locally supersymmetric theories.
 It would be very interesting to learn more about the structure of these theories. 
 Their (effective) description is not in terms of a Lagrangian but only in terms of their operator content, their massive
 fields and their on-shell scattering amplitudes.
 Hence relevant information might be obtained by computing string scattering amplitudes with massive external fields, as it was done 
 e.g. in  \cite{Feng:2010yx,Feng:2012bb}.

Finally we like to comment about a possible relation between the holographic picture and the double copy picture of supergravity theories.
This follows from 
the intriguing correspondence between the 4D ${\cal N}$-extended Weyl multiplets
and the 5D standard $2{\cal N}$-extended supergravity multiplets. This is because they both deal with 4D massive spin-2 respresentations of the same superconformal
algebra. In this way, the  4D superconformal algebra of ${\cal N}$-extended quantum field theories (${\cal N}\leq4$) at the four-dimensional boundary is holographically equivalent
to the  supersymmetry algebra of $2{\cal N}$-extended $AdS_5$ supergravity theories. However the difference between this two cases is that for the double
copy construction the spin-two field lives on the 4D boundary, whereas in the holographic duality the spin-two fields lives in the 5D bulk space.
Following this observation, if we identify Weyl supergravity with the superconformal ${\cal N}=4$ boundary theory, then its holographic dual should be an ${\cal N}=8$ a spin-four
W-supergravity theory (for comprehensive  reviews and some papers on higher spin theories see    \cite{Bekaert:2005vh,Alday:2007mf,Henneaux:2010xg,Sagnotti:2013bha}).
Note that four-dimensional Weyl supergravity violates unitarity do to the quartic derivative action which leads to a dipole ghost \cite{Stelle:1976gc,Alvarez-Gaume:2015rwa}.
It is therefore an interesting question if this also implies a lack of unitarity in the spin-four W supergravity.

\section*{Acknowledgements:}

We like to thank  Inaki Garcia-Etxebarria, Wolfgang Lerche, Diego Regalado and in particular Ioannis Florakis for very useful discussions. The work of S.F. is supported in part
by CERN TH Dept and INFN-CSN4-GSS. The work of D.L. is supported by the ERC Advanced Grant ``Strings and Gravity" (Grant No. 320045) and the Excellence Cluster Universe.
He also is grateful to the CERN theory department for its hospitality, when part of this work was performed.

\section*{Appendix: Supercharges}

Starting from ten dimensions, the 16 supercharges $Q^{\alpha, A}$ and $Q^{\dot \alpha ,\dot A}$ belong to the 16-dimensional spinor representation ${\underline {16}}_s$ of the group $SO(10)$.
Decomposing this group into the four-dimensional Lorentz-group times the R-symmetry group, i.e. $SO(10)\supset SO(1,3)\times SO(6)_R$, the spinor ${\underline {16}}_s$
decomposes as
\begin{equation}
Q^{\alpha, A}\, , Q^{\dot \alpha, \dot A}:\quad {\underline {16}}_s=({\underline 2}_s,{\underline 4}_s)+({\underline 2}_c,{\underline 4}_c)\, .
\end{equation}
Further we also need  the 10-dimensional transversal Lorentz-group $SO(8)_T$, which is the little group for massless states, and under which 
the spinor  of $SO(10)\supset SO(8)_T\times SO(2)_L$ decomposes as
\begin{equation}
{\underline {16}}_s=({\underline 8}_s,{1\over 2})+({\underline 8}_c,-{1\over 2})\, 
.
\end{equation}
Here $SO(2)_L$ is the rotation group, which acts on the longitudinal degrees of freedom of the massless states.

Finally we need also the branching of $SO(8)_T$ into the four-dimensional,
transversal little group $SO(2)_T$ times the R-symmetry group $SO(6)_R$, i.e. $SO(8)_T\supset SO(2)_T\times SO(6)_R$:
\begin{equation}
{\underline {8}}_s=({1\over 2},{\underline 4}_s)+(-{1\over 2},{\underline 4}_c)\, 
,\quad \, {\underline {8}}_c=(-{1\over 2},{\underline 4}_s)+({1\over 2},{\underline 4}_c)\,
.
\end{equation}
Comparing the different decompositions, one sees that the eight supercharges $Q^{\alpha ,A}$ belong to the two first representations in ${\underline {8}}_s$ and ${\underline {8}}_c$,
namely
\begin{equation}
Q^{\alpha ,A}:\quad     ({1\over 2},{\underline 4}_s)+  (-{1\over 2},{\underline 4}_s) \, ,
\end{equation}
whereas the other eight 
supercharges $Q^{\dot\alpha ,\dot A}$ belong to the two second representations in ${\underline {8}}_s$ and ${\underline {8}}_c$,
namely
\begin{equation}
Q^{\dot\alpha ,\dot A}:\quad     ({1\over 2},{\underline 4}_c)+  (-{1\over 2},{\underline 4}_c) \, ,
\end{equation}

\end{document}